\documentclass[10pt,conference]{IEEEtran}
\usepackage[hidelinks]{hyperref}

\usepackage[nocompress]{cite}
\usepackage[T1]{fontenc}
\usepackage{amsmath,amssymb,amsfonts}
\usepackage{algorithmic}
\usepackage{array}
\usepackage[caption=false,font=footnotesize]{subfig}
\usepackage{textcomp}
\usepackage{afterpage}
\usepackage{stfloats}
\usepackage{url}
\usepackage{verbatim}
\usepackage[normalem]{ulem}
\def\BibTeX{{\rm B\kern-.05em{\sc i\kern-.025em b}\kern-.08em
    T\kern-.1667em\lower.7ex\hbox{E}\kern-.125emX}}
\usepackage{balance}

\usepackage{graphicx}
\graphicspath{{./Figs/}}
\DeclareGraphicsExtensions{.pdf,.jpeg,.png,.eps}
\usepackage{adjustbox}
\usepackage{booktabs}
\usepackage{xurl}  
\usepackage{multirow}
\usepackage{hhline}
\usepackage{tabularx}
\usepackage{siunitx} 
\usepackage{xspace}
\usepackage[table]{xcolor}
\usepackage{hyperref}
\usepackage{pbox} 

\usepackage{seqsplit}
\usepackage{csquotes}

\usepackage{tabularx}
\usepackage{enumerate}
\usepackage[most]{tcolorbox}
\usepackage{float}
\usepackage{tikz}
\usepackage{array}
\usepackage{makecell}
\usepackage{booktabs}
\usepackage{tabularx, array, makecell, threeparttable} 
\newcolumntype{L}[1]{>{\raggedright\arraybackslash}m{#1}} 
\newcolumntype{R}[1]{>{\raggedleft\arraybackslash}m{#1}}  
\newcolumntype{Y}{>{\raggedright\arraybackslash}X}        

\usepackage{booktabs}      

\definecolor{darkgreen}{RGB}{0,128,0}
\definecolor{darkred}{RGB}{178,34,34}
\definecolor{gray}{RGB}{128,128,128}

\definecolor{rank1}{HTML}{99FF99}
\definecolor{rank2}{HTML}{FFFF99}
\definecolor{rank3}{HTML}{FFCC99}
\definecolor{rank4}{HTML}{FF9999}
\definecolor{rank5}{HTML}{D4A0A0}
\definecolor{rank6}{HTML}{C090C0}

\colorlet{dgreen}{darkgreen}
\colorlet{dred}{darkred}

\newtcolorbox{Summary}[3][]{
    enhanced,
    sharp corners,
    colback=white,
    colframe=black,
    boxrule=0.4pt,
    left=4pt,
    right=4pt,
    top=3pt,
    bottom=3pt,
    boxsep=0pt,
    before skip=4pt,
    after skip=4pt,
    fontupper=\normalsize,
    #1,
    before upper={\textbf{#2.}\ }
}

\definecolor{custom-gray}{cmyk}{0, 0, 0, 0.7, 1.00}

\title{Developer-LLM Conversations: An Empirical Study of Interactions and Generated Code Quality}

\author{
    \IEEEauthorblockN{Suzhen Zhong}
    \IEEEauthorblockA{
    Queen's University Canada\\
    suzhen.zhong@queensu.ca
    }
    \and
    \IEEEauthorblockN{Ying Zou}
    \IEEEauthorblockA{
    Queen's University Canada\\
    ying.zou@queensu.ca
    }
    \and
    \IEEEauthorblockN{Bram Adams}
    \IEEEauthorblockA{
    Queen's University Canada\\
    bram.adams@queensu.ca
    }
}

\newcommand{\dataset}{CodeChat\xspace}

\newcommand{\rqtwo}{What topics do developers most commonly prompt when interacting with LLMs?}
\newcommand{\rqthree}{How high is the quality of LLM-generated code within coding conversations?}

\newcommand{\motivation}{\textbf{\textit{Motivation. }}}

\newcommand{\findings}{\textbf{\textit{Results. }}}

\begin{document}
\maketitle
\begin{abstract}
Large Language Models~(LLMs) are increasingly used by developers for conversational coding assistance, including code generation, technical questions, and iterative problem solving. However, limited empirical evidence exists on the topics developers discuss with LLMs, how interaction patterns vary across tasks, and how the quality of generated code changes across conversational turns. To address this, we construct CodeChat~\cite{codechat_dataset}, a large-scale dataset derived from the WildChat dataset~\cite{zhao2024wildchat}. CodeChat contains 587,568 real-world developer-LLM conversations, containing 1.7 million code snippets across more than 20 programming languages. We identify the topics developers raise and examine how their engagement varies across topics. We then assess the quality of generated code across five programming languages~(i.e., Python, JavaScript, C++, Java, and C\#) and analyze how quality issues evolve across conversational turns. We find that developers most frequently seek help with web design (9.6\% of conversations) and machine learning (8.7\%), and that conversations often grow long because developers repeatedly shift use cases. LLM-generated code exhibits prevalent, language-specific issues whose
prevalence does not consistently decrease in interactions lasting at least
five turns. Among successful syntax-error resolutions, prompts that explicitly identify the error and request a fix are the most frequently observed. Our findings suggest that conversational assistants should track evolving developer intent and monitor code quality across turns.
\end{abstract}
\begin{IEEEkeywords}
Developer-LLM Conversations, LLM Code Generation, Code Quality, Vibe Coding, AI-Assisted Programming
\end{IEEEkeywords}

\section{Introduction}\label{sec:introduction}

Developers have long relied on programming question-and-answer~(Q\&A) platforms such as Stack Overflow to seek help from peers and resolve technical problems~\cite{Barua2014,so_hci_humancode}. More recently, large language models~(LLMs), such as ChatGPT~\cite{chatgpt2025} and Claude~\cite{claude2025}, have introduced a new way for developers to obtain programming support. Trained on large corpora of text and code~\cite{NIPS2017_3f5ee243}, these models can respond directly to natural-language requests and support iterative, conversational problem solving. Therefore, LLM-based assistants have emerged as an alternative to traditional Q\&A platforms such as Stack Overflow~\cite{dasilva2024chatgptvsllamaimpact}. Developers now use them for a range of software engineering tasks, including code generation~\cite{jiang2024surveylargelanguagemodels,10.1145/3697012} and code quality evaluation~\cite{10.1145/3701625.3701650,haider2024promptingfinetuninglargelanguage}.
 
\begin{figure}[t]
  \centering
  \begin{minipage}{0.49\columnwidth}\centering
    \includegraphics[width=\linewidth,trim=8 8 8 8,clip]{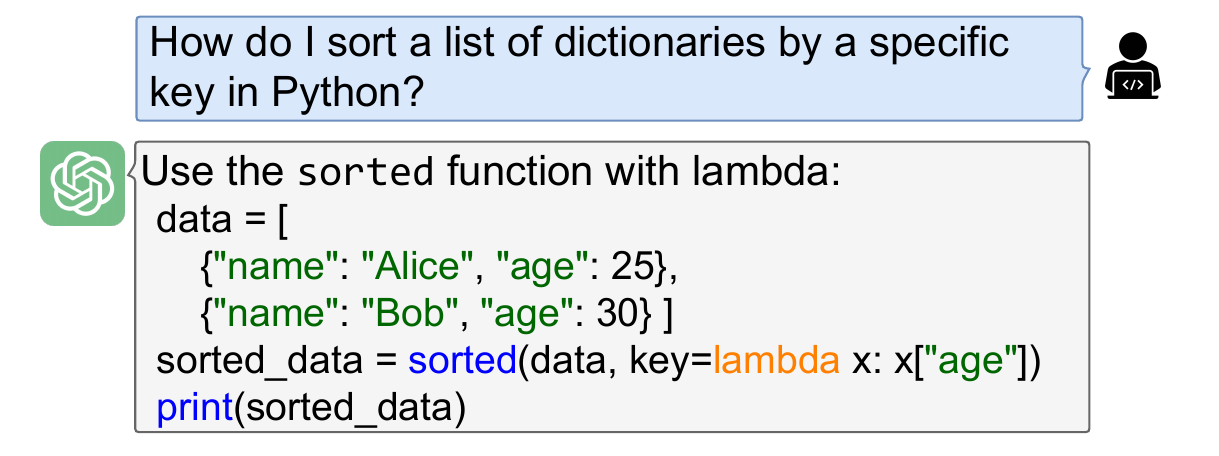}
    \scriptsize (a) Single-turn
  \end{minipage}\hfill
  \begin{minipage}{0.49\columnwidth}\centering
    \includegraphics[width=\linewidth,trim=8 8 8 8,clip]{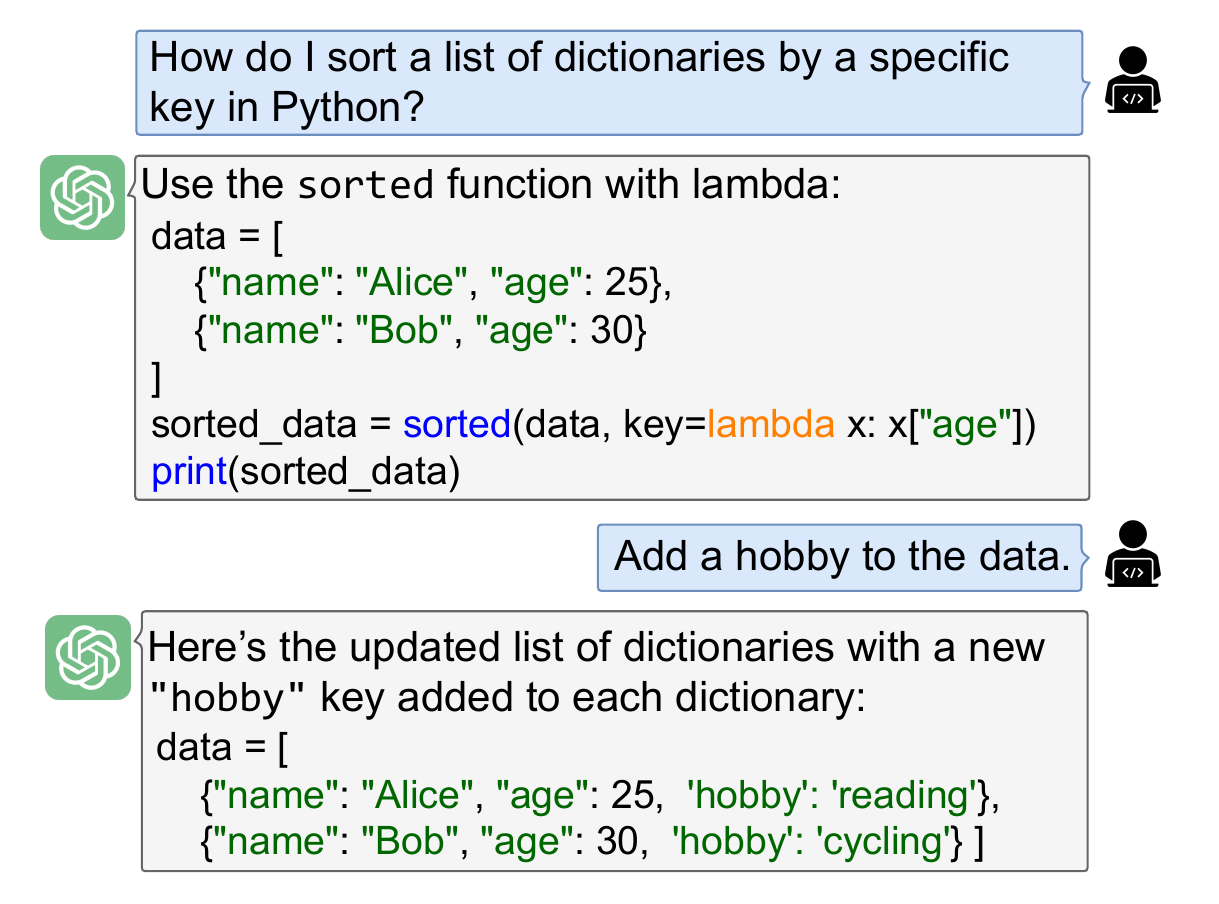}
    \scriptsize (b) Multi-turn
  \end{minipage}
  \caption{Examples of developer–LLM conversations.}
  \label{fig:developer-LLM-conversations}
  \vspace{-15pt}
\end{figure}

As illustrated in Figure~\ref{fig:developer-LLM-conversations}, conversations between developers and LLMs may consist of either a single prompt-response pair or multiple turns. In a single-turn conversation, the developer submits an isolated request and receives one response. In a multi-turn conversation, the developer follows up on earlier responses, refines the request, or provides additional context. Across these interactions, developers may ask technical questions, brainstorm solutions, request code, or seek explanations and feedback.

However, detailed conversational data from commercial LLM services such as OpenAI and Anthropic remains unavailable to the public due to privacy concerns~\cite{openai2025privacy} and intellectual property protections~\cite{anthropicPrivacy2025}. This limitation restricts comprehensive analyses and empirical studies. To address the lack of publicly accessible data, prior research has mined explicitly shared ChatGPT conversations from GitHub and Hacker News~\cite{xiao2024devgpt}. More recent work has built datasets of repository artifacts, such as commits, issues, and pull requests, produced by autonomous coding agents~\cite{li2025riseai,baumann2026swechat}. In contrast, developers' daily use of LLMs often consists of quick, interactive back-and-forth conversations that occur outside a repository, which existing datasets do not capture.

In this study, we introduce \dataset~\cite{codechat_dataset}, a large-scale dataset of real-world developer-LLM conversations obtained by filtering the WildChat dataset~\cite{zhao2024wildchat}, which comprises interactions mined from publicly accessible chatbot services. \dataset comprises 587,568 developer-LLM conversations, totalling 1.1 million turns and 1.7 million code snippets across more than 20 programming languages. Using \dataset, we study the common topics developers raise and the quality of the code that LLMs generate, through the following research questions~(RQs):

\textbf{RQ1. \rqtwo} Analyzing common topics in developer-LLM interactions reveals how developers use LLMs across different tasks and where certain use cases require more extensive interaction. We apply BERTopic~\cite{grootendorst2022bertopic}, a transformer-based topic modeling technique, to \dataset prompts and identify the most frequent topics. The most prevalent topics are web design (9.6\% of conversations) and machine learning (8.7\%). Developer–LLM engagement varies by topic, with AI-augmented business tool automation showing the highest interaction~(mean = 3.40 turns). Within the topic with the highest mean turn count, repeated use-case shifts, feature revisions, and unresolved erroneous responses appear among the frequent prompt-gap patterns.

\textbf{RQ2. \rqthree} Conversational LLMs remain widely used for quick coding assistance, even as more autonomous coding agents become available. Evaluating generated code quality shows whether short exchanges provide reliable programming support and whether additional turns fix earlier defects. Across five major languages (i.e., Python, JavaScript, C++, Java, and C\#), we find widespread defects in first-turn LLM-generated code, particularly undefined variables and missing documentation. In multi-turn conversations, code quality does not consistently improve. For example, undefined-variable errors in Python increase from 23.5\% to 32.8\% after five turns, whereas documentation violations in Java decrease from 78.1\% to 63.4\%. Users often correct syntax errors by identifying the error, asking targeted questions, or providing specific instructions. These findings show that additional turns alone do not guarantee better code quality, while explicit and focused feedback is commonly associated with successful corrections.

The main contributions of this paper are as follows:

\begin{enumerate}[1.]
    \item We provide \dataset~\cite{codechat_dataset}, a comprehensive dataset of real-world developer-LLM conversations obtained from WildChat~\cite{zhao2024wildchat}.
    \item We characterize the topics of developer-LLM conversations and examine the duration of the interaction, showing which tasks involve more extensive conversational support.   
    \item We conduct a large-scale, multi-language analysis of LLM-generated code quality across multi-turn conversations, showing that additional interaction does not consistently improve code quality.
\end{enumerate}

To support reproducibility, we release the \dataset dataset and make our code and analysis scripts available in a public replication package~\cite{sealXYZ2025}.



\section{Experiment Setup}\label{sec:methodology}

\begin{figure*}[t]
	\centering
	\includegraphics[width=\textwidth]{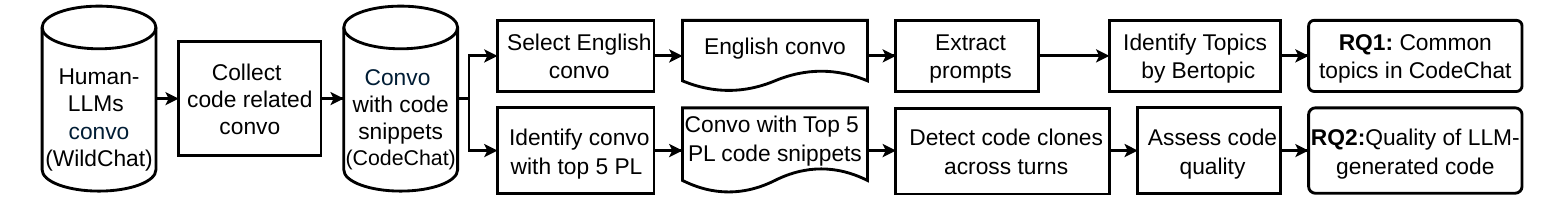}
    \vspace{-20pt}
	\caption{Overview of the approach. The term ``convo'' refers to ``conversations'', and ``PL'' stands for ``Programming Language''.}
	\label{fig:methodology}
    \vspace{-15pt}
\end{figure*}

Figure~\ref{fig:methodology} shows an overview of our approach. We derive the \dataset dataset from WildChat~\cite{zhao2024wildchat} and use it to answer two research questions about the common topics developers ask LLMs about~(RQ1) and LLM-generated code quality~(RQ2). 


\subsection{Data Collection}\label{sec:methodology_1}
The dataset that we use is derived from the WildChat~\cite{zhao2024wildchat}, a large-scale collection of 3.2 million real-world human–LLM conversations comprising 5.2 million conversational turns. Collected from April 2023 to July 2025 across 76 languages and 1.8 million users, it reflects recent and diverse ChatGPT usage and is publicly available on Hugging Face. Compared to LMSYS-Chat-1M~\cite{zheng2023lmsys}, which includes about 70\% Vicuna~\cite{vicuna2023} interactions, WildChat is more relevant due to its focus on ChatGPT, a widely adopted model. In addition, OpenAssistant~\cite{kopf2024openassistant} ends data collection in April 2023 and contains 66,479 conversations, making WildChat~(with 3.2 million conversations) both more current and substantially larger.

Conversations in WildChat are represented in a JSON format to support systematic analysis and reproducibility. A WildChat entry contains metadata such as \texttt{conversation\_id} and \texttt{model} name, along with a \texttt{conversation} field containing a list of dialogue turns. Each turn includes attributes such as \texttt{role} (e.g., user or LLM), \texttt{content}, \texttt{language}, \texttt{timestamp}, and \texttt{hashed\_ip}. 

From WildChat, we identify code-related conversations by examining the \texttt{content} field of each LLM turn. We consider a conversation code-related if at least one LLM turn contains a code block enclosed in triple backticks~(\enquote{\`{}\`{}\`{}}) followed by a programming language identifier, such as \texttt{Python}. This criterion captures conversations in which the LLM explicitly generates programming code while enabling us to identify the corresponding programming language. Applying this filter yields 587,568 conversations (18.4\% of WildChat), comprising 1.1 million conversational turns and 1.7 million code snippets. We refer to the resulting dataset as \dataset. \dataset contains interactions associated with 426,062 unique hashed IP addresses, which we use as a proxy for users. To clarify participant roles, we refer to the \texttt{user} in these code-related conversations as the \texttt{developer}. All other metadata is inherited directly from WildChat without modification.

\subsection{Dataset Analysis}\label{sec:methodology_2}
We characterize \dataset by comparing it to existing datasets and by analyzing its conversation turns and programming languages.

Table~\ref{tab:rw_dataset} compares \dataset with two existing developer–LLM conversation datasets, DevGPT~\cite{xiao2024devgpt} and GPT-Social~\cite{10196869}. \dataset is substantially larger, with 587,568 conversations and 1.7 million LLM-generated code snippets, compared to 29,778 conversations in DevGPT and 332 in GPT-Social. It also spans a longer and more recent period of 28 months, from April 2023 to July 2025, and covers more models, from GPT-3.5 to o1. Together, these properties make \dataset a larger, more recent, and more diverse basis for studying developer–LLM interactions.

\begin{table*}[!t]
\caption{Overview of conversational datasets.}
\label{tab:rw_dataset}
\centering
\begin{tabular}{@{}llrrclll@{}}
\toprule
                    & \textbf{Source} & \textbf{\#Conversation} & \textbf{\#Snippets} & \textbf{Platform}                                              & \textbf{\#Users} & \textbf{Model(s)} & \textbf{\begin{tabular}[c]{@{}c@{}}Time\\ (Duration)\end{tabular}} \\ \midrule
\textbf{DevGPT}     & \cite{xiao2024devgpt} & 29,778           & 19,106              & \begin{tabular}[c]{@{}c@{}}GitHub, \\ Hacker News\end{tabular} & -                & GPT-3.5           & \begin{tabular}[c]{@{}c@{}}2023.07 -- 2023.10\\ (4 months)\end{tabular} \\
\midrule
\textbf{GPT-Social} & \cite{10196869} & 332              & -                   & \begin{tabular}[c]{@{}c@{}}Twitter/X, \\ Reddit\end{tabular}       & -                & Not reported      & \begin{tabular}[c]{@{}c@{}}2023.12 -- 2024.01\\ (2 months)\end{tabular} \\
\midrule
\textbf{\dataset}   & This study & 587,568          & 1,724,902           & HuggingFace                                                              & 426,062$^*$      & \begin{tabular}[c]{@{}l@{}}
GPT-3.5; GPT-4o; \\
GPT-4; GPT-4.1;o1;
\end{tabular} & \begin{tabular}[c]{@{}c@{}}2023.04 -- 2025.07\\ (28 months)\end{tabular} \\
\bottomrule
\multicolumn{8}{@{}l@{}}{\scriptsize $^{*}$ Estimated based on the number of unique IP addresses.} \\
\end{tabular}
\vspace{-10pt}
\end{table*}

\begin{figure}[t]
	\centering
    \includegraphics[width=\columnwidth]{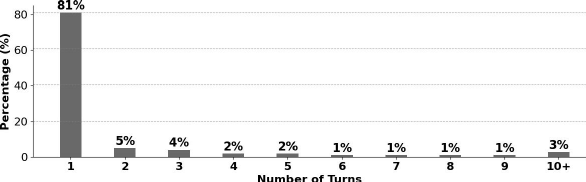}
    \vspace{-20pt}
	\caption{Turns Distribution in \dataset}
	\label{fig:RQ0_turns}
    \vspace{-10pt}
\end{figure}
\begin{figure}[t]
    \centering
    \includegraphics[width=\columnwidth]{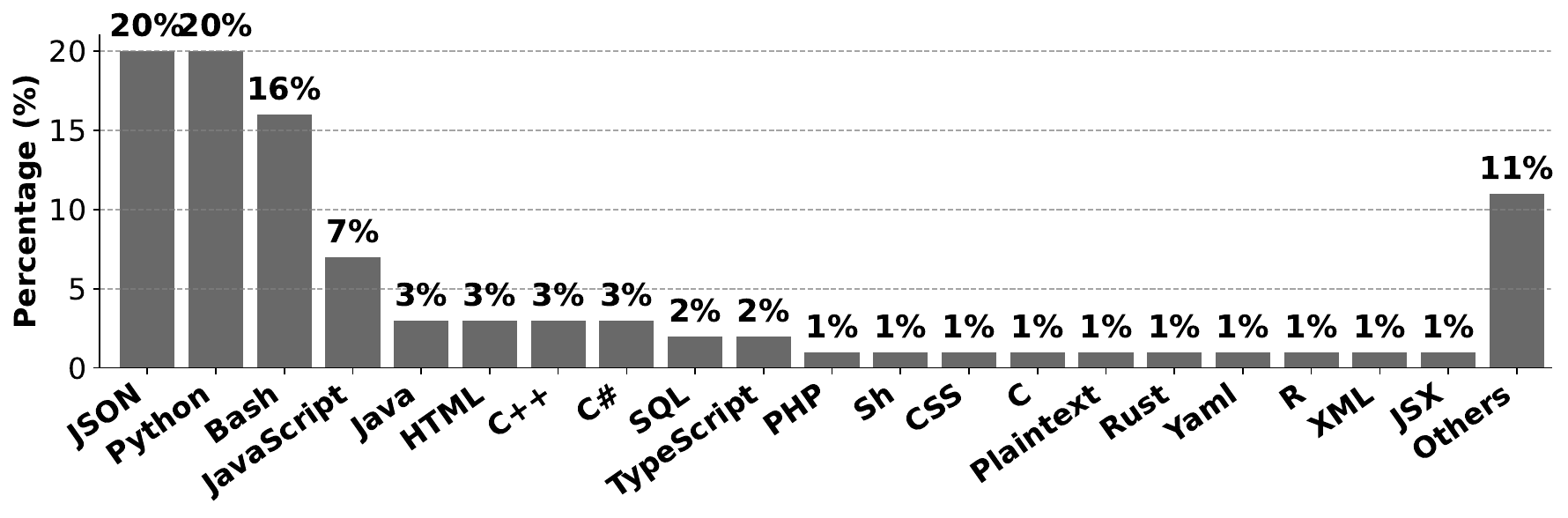}
    \vspace{-15pt}
    \caption{Top 20 programming languages in \dataset.}
    \label{fig:RQ0_lang_loc}
    \vspace{-12pt}
\end{figure}

Figure~\ref{fig:RQ0_turns} shows the distribution of conversation length, measured by the number of prompt–response pairs~(turns) in each conversation. Most \dataset conversations are single-turn (81\%), consisting of one developer prompt and one LLM response. The remaining 19\% are multi-turn. Although most conversations have fewer than 10 turns, the distribution has a long tail, with 3\%~(15,409 conversations) extending to 10 turns or more. Figure~\ref{fig:RQ0_lang_loc} shows the top 20 programming languages generated in \dataset. The most frequent are JSON (20\%), Python (20\%), and Bash (16\%), followed by JavaScript (7\%), Java, and HTML (3\% each).

\subsection{Common Topics in CodeChat}
To characterize developer–LLM interactions, we use BERTopic~\cite{grootendorst2022bertopic} to identify common topics from developers' initial prompts and examine how developers engage in back-and-forth interactions within each topic. We further investigate why some topics lead to more prolonged conversations.

\textbf{Select English Conversations.} To identify the topics that initiate developer--LLM conversations, we extract the first developer turn from each conversation. We restrict topic modeling to English prompts because BERTopic's default embedding model is optimized for English text~\cite{grootendorst2022bertopic}. Using the language metadata inherited from WildChat, we retain English conversations and use their initial prompts as input for topic modeling.

\textbf{Identify common topics.} We apply BERTopic~\cite{grootendorst2022bertopic}, a transformer-based topic modeling technique widely used in software engineering research~\cite{Kumar2025,Gottumukkala2025}, to group developers' initial prompts into topics. To identify the optimal hyperparameters, we conduct a grid search guided by the default and recommended settings from the BERTopic documentation~\cite{Gottumukkala2025}. We then select the optimal clustering configuration using a combination of quantitative metrics and qualitative interpretability. From the grid search results, we select the top 10 hyperparameter settings based on topic coherence scores~\cite{10.1145/2684822.2685324}, which evaluate the semantic interpretability and consistency of the generated topics by measuring the co-occurrence probabilities of the top words within each topic.

To select the best-performing configuration, we manually examine the representative prompts generated by BERTopic for each of the top 10 configurations. Specifically, we manually assess whether representative prompts in each group share a coherent theme and align with the topic’s top keywords, discarding configurations with broad or inconsistent topics. We then retain one configuration that produces the most coherent and meaningful topics for further analysis.

To examine whether developers stay on their initial topic or drift to new ones, we track how their intent evolves across multi-turn conversations. For each conversation, we assign an initial topic from the developer's first prompt. For each subsequent turn, we apply the trained BERTopic model to the cumulative prompts to infer the topic. We define an intent shift as a drop of more than 5\% in the initial topic's probability, and mark the transition to the new topic. This assigns a topic to every turn, which the following analyses use to measure engagement and language use per topic.

\begin{table}[t]
\centering
\caption{Summary of prompt design gap types~(Adapted from \cite{10.1145/3643991.3645085})}
\label{tab:rq1_prompt_gaps}
\begin{tabularx}{\columnwidth}{@{}c @{\hspace{6pt}} X @{}}
\toprule
\textbf{Prompt Design Gap Type} & \textbf{Description} \\ \midrule
Missing Specifications & Lacks explicit requirements or key information. \\ \midrule
Different Use Cases & Addresses different problems within one conversation. \\ \midrule
Incremental Problem Solving & Breaks the task into smaller components, solved step-by-step. \\ \midrule
Exploring Alternative Approach & Requests alternative solutions to a problem. \\ \midrule
Wordy Response & Seeks to reduce unnecessary verbosity in the LLM’s output. \\ \midrule
Additional Functionality & Requests new features or extensions to an existing solution. \\ \midrule
Erroneous Response & Addresses errors in previous LLM output. \\ \midrule
Missing Context & Lacks sufficient background for the LLM to generate an appropriate response. \\ \midrule
Clarity of Generated Response & Aims to clarify ambiguous responses. \\ \midrule
Untrustworthy Response & Responses fail to operate as intended or generated outputs lack reliability. \\ \midrule
Miscellaneous & Prompts not fitting other categories (e.g., gratitude, null prompts). \\
\bottomrule
\end{tabularx}
\vspace{-12pt}
\end{table}


\textbf{Analyze engagement patterns across topics.} To understand how developers engage with different topics, we compute the turn count (TC) for each conversation within each identified topic. A higher TC for a topic indicates more extensive interactions, which may reflect increased task complexity or the need for iterative clarification. We then employ the Scott–Knott test~\cite{scott1974cluster,zhong2026humanaisynergyagenticcode,zhong2026humancentricagenticcodereview}, which recursively partitions topics into clusters with similar mean turn counts (TC), to group topics with similar engagement levels. Unlike pairwise statistical tests, Scott–Knott considers all topics simultaneously and produces an interpretable partitioning of engagement levels across all topics. The resulting clusters identify topics associated with higher or lower engagement levels, as measured by the number of conversational turns.

\textbf{Investigate the causes of prolonged interactions.} To examine why some topics lead to prolonged interactions, we examine prompt design gaps~(i.e., the reasons a developer submits a follow-up prompt within a conversation~\cite{10.1145/3643991.3645085}) as described in Table~\ref{tab:rq1_prompt_gaps}. Within a multi-turn conversation, we label each consecutive developer prompt pair with its gap type, producing a sequence of gap labels that we call a \textit{chain-of-gap pattern}. Focusing on the topic cluster with the highest average turn count (as determined by the Scott–Knott test), we randomly sample and manually label 383 conversations (95\% confidence level, 5\% margin of error), resulting in a gap chain for each conversation. To identify recurring prompt design patterns in prolonged interactions, we extract all consecutive 3-grams of gap types from each gap chain~(allowing repeats) and compute their prevalence.

\subsection{Quality of LLM generated Code}
To assess the quality of LLM-generated code and how it evolves across multi-turn interactions, we select the five most frequently generated languages that support robust static code analysis: Python, JavaScript, C++, Java, and C\#. We exclude JSON and HTML (the 1st and 6th most frequent) because they contain no executable logic to assess. We also exclude shell scripts such as Bash, which have limited structural complexity and weak tooling.

\textbf{Detect code clones in multi-turn conversations.} From the multi-turn conversations in the five selected languages, we analyze how code quality issues evolve across the outputs of consecutive turns within the same conversation. As unrelated tasks are common in multi-turn conversations (Section \ref{sec:RQ2}), we filter out subsequent turns where developers switch tasks or the LLM generates unrelated code. To detect such task switches, we apply code clone detection by checking the similarity between code snippets from two adjacent turns. If two successive conversation turns yield sufficiently similar code, we consider both turns to be related and add the second turn to the current task sequence. A \textbf{task sequence} is defined as a consecutive set of turns where the generated code remains sufficiently similar, reflecting work on the same programming task. However, if the similarity between both turns' code is too low, we consider that a new task has started, ending the previous task sequence.

We employ C4~\cite{10.1145/3524610.3527911}, a contrastive learning-based model for detecting near-miss~(i.e., code snippets that maintain structural similarity despite modifications such as statement addition, removal, or alteration) and semantic code clones~(i.e., code snippets that achieve the same functional behaviour despite different syntactic or structural implementations) across programming languages. C4 compares learned code representations to identify consecutive outputs that are structurally or functionally related, indicating contextual continuity rather than a shift to a new or unrelated prompt. 

\textbf{Assess the evolution of code quality.\label{RQ3_evolution} } To evaluate the evolution of LLM-generated code quality, we use established code quality tools to detect potential defects~(e.g., undefined variables, null pointer dereferences) and code smells~(e.g., stylistic issues) across the different code versions in the analyzed task sequences. For Python and Java, we follow the approach in ~\cite{liu2024refining}, which leverages Pylint~\cite{pylint2025} for Python and PMD~\cite{pmd2025} for Java. For JavaScript, we use ESLint~\cite{eslint2025}, which is a widely adopted tool for detecting syntax errors and style violations. For C++, we employ Cppcheck~\cite{cppcheck2025}, which captures a broad range of issues, including performance and portability concerns. For C\#, we utilize Roslyn~\cite{roslynAnalyzers2025}, which is integrated into the .NET SDK for inspecting syntax, semantics, and maintainability issues.

For each programming language, we identify the most frequent quality issue types detected in the first-turn response and track their prevalence across Turns~1--5. To ensure that issue prevalence is compared across the same population, we restrict this longitudinal analysis to task sequences containing at least five turns. We compute the percentage of retained task sequences containing at least one instance of quality issue type $m$ at turn $N$ as $P_{m,N} = \frac{C_{m,N}}{C_5} \times 100$, where $C_{m,N}$ is the number of retained task sequences containing at least one occurrence of issue type $m$ at turn $N$, and $C_5$ is the total number of task sequences containing at least five turns.

To evaluate whether the frequency of specific quality issue types changes across conversational turns, we conduct trend analysis on the computed $P_{m,N}$ values for each issue type. Specifically, we perform linear regression\cite{montgomery2021introduction}, modeling $P_{m,N}$ as a function of turn number $N$. A statistically significant regression slope ($p < 0.05$) is interpreted as evidence of a trend. In cases where the data may violate linearity or normality assumptions, we apply nonparametric Mann-Kendall trend test\cite{mannkendall1998}, which is robust to non-normal distributions and outliers. This combined methodology ensures the reliability and validity of our assessment of issue prevalence trends across multi-turn interactions.

\textbf{Analyze prompts that resolve syntax errors.} To characterize follow-up prompts preceding syntax-error-free responses, we analyze adjacent response pairs in multi-turn conversations where the first response contains at least one detected syntax error and the second contains none. The syntax error quality issue type is chosen because it represents fundamental correctness issues that prevent code execution and can be reliably detected by static analysis. From \dataset, we extract multi-turn conversations with Python, JavaScript, or C++ code snippets that include at least one pair of adjacent LLM responses in which the first contains at least one syntax error and the second contains none. To ensure statistical validity, we randomly sample 383 cases from these conversations using the standard estimation method for large populations at a 95\% confidence level and a 5\% margin of error.

Labeling is performed independently by the first author and a fifth‑year Ph.D. student in software engineering, following \cite{11025620}'s nine‑category scheme~(as shown in Table~\ref{tab:syntax_error_reduction_categories}) of conversational prompts for guiding LLMs in code generation tasks. We inductively define two additional categories: \textit{Change Use Cases}, where the developer redirects the LLM to a different coding goal, and \textit{Request Full Code}, where the developer asks for the complete corrected program rather than partial fixes. Inter‑annotator agreement for all categories is computed using Cohen’s kappa and equals 0.90, indicating almost perfect agreement.

\section{Results}\label{sec:results}


\subsection{\textbf{RQ1: \rqtwo}}\label{sec:RQ2}

\motivation
Developers interact with LLMs to address a wide range of coding tasks, reflecting diverse intentions and tasks in their daily workflows. However, the specific topics developers prompt in actual practice remain largely unexplored. Understanding these topics is crucial for identifying the most pressing needs of developers and ensuring that LLMs provide relevant and effective support. In this RQ, we aim to uncover and categorize the common topics that developers prompt. These insights will help code assistant developers and researchers improve LLMs to better align with real-world programming tasks.

\begin{figure}[t]
    \centering\includegraphics[width=\columnwidth]{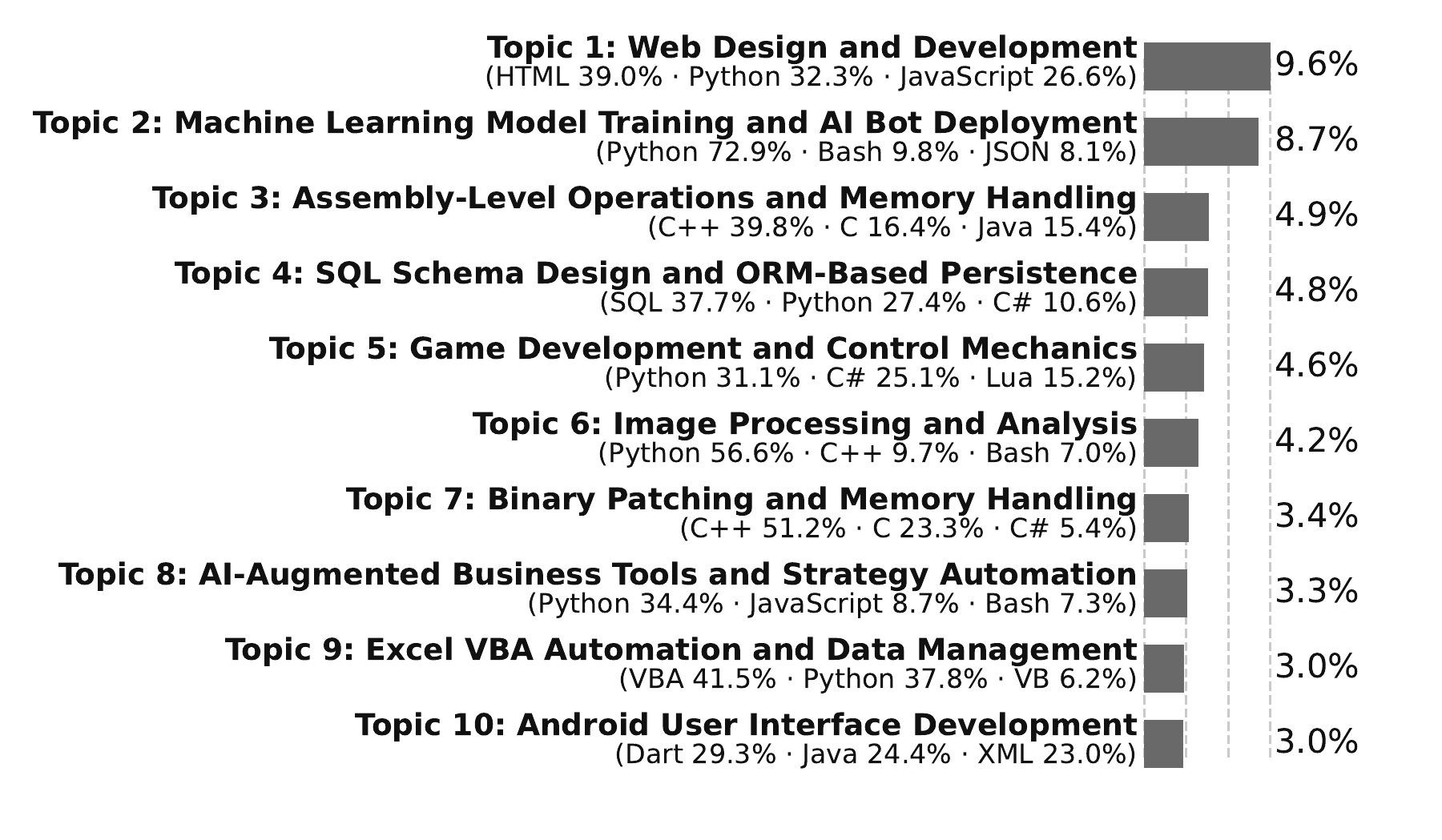}
    \medskip\includegraphics[width=\columnwidth]{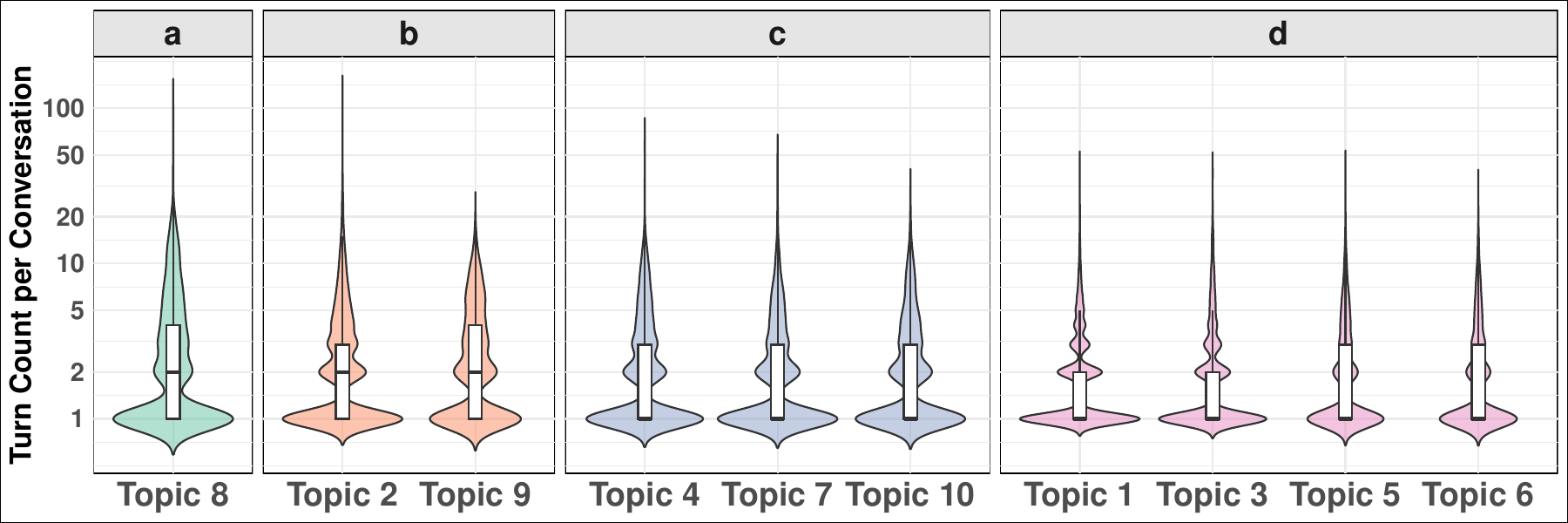}
    \vspace{-25pt}
    \caption{(a) Top 10 most frequent topics in developer–LLM conversations; 
    (b) Scott–Knott groupings for turn-count distributions per topic, reflecting the number of interactions across topics.}
    \label{fig:RQ2_topics_and_turns}
    \vspace{-10pt}
\end{figure}

\findings\textbf{Web design and development is the most frequent topic, comprising 9.6\% of conversations.} As shown in Figure~\ref{fig:RQ2_topics_and_turns}, developers seek assistance with front-end tasks such as layout design, responsive styling, and interactive components. This aligns with Anthropic’s analysis~\cite{anthropic2025}, which reports that coders commonly use AI for building user-facing applications, especially with web-development languages such as JavaScript and HTML, and that user interface tasks are among the top coding uses. Common keywords in this topic include \texttt{div}, used for structuring web pages in HTML~\cite{mdnDiv2025}, and \texttt{react}, a JavaScript library for building user interfaces~\cite{react2025}.

\textbf{Machine learning model training and AI bot deployment is the second most frequent topic, comprising 8.7\% of conversations.} This topic includes prompts where developers seek assistance with configuring training pipelines, such as implementing gradient scaling in PyTorch~\cite{paszke2019pytorch}, designing deep learning architectures in Keras~\cite{chollet2015keras}, and optimizing multi-task learning workflows~\cite{crawshaw2020multi}. Prompts also address challenges related to deploying AI models, including inference request handling and output parsing when integrating model APIs.

\textbf{Low-level programming and memory manipulation accounts for 8.3\% of conversations, spanning assembly-level operations~(4.9\%) and binary patching (3.4\%).} The predominant languages in these topics are \texttt{C++} and \texttt{C}, reflecting their central role in systems-level development. Developer prompts involve fine-grained memory control and low-level system behavior, with assembly instructions such as \texttt{mov}, \texttt{ptr}, and \texttt{eax}. In binary patching, developers seek assistance with runtime code modification, driver-level memory operations, and byte-level injection. These interactions highlight the application of LLMs in supporting tasks related to systems programming, reverse engineering, and low-level debugging.

\textbf{Developer–LLM engagement differs significantly across topics as measured by TC, with four statistically distinct groups.} As shown in Figure~\ref{fig:RQ2_topics_and_turns}, Topic 8 (AI-augmented business tools and strategy automation) on its own forms one group~(group ``a'') identified by the Scott-Knott analysis, and shows the highest engagement with a mean of 3.4 turns. Group b includes Topic 2~(machine learning model training and AI bot deployment, mean = 2.89) and Topic 9~(Excel VBA automation and data management, mean = 2.86), both showing moderately high engagement. Group ``c'' contains Topic 4~(SQL schema design and ORM-based persistence, mean = 2.62), Topic 10~(Android user interface development, mean = 2.50), and Topic 7~(Binary patching and memory handling, mean = 2.42), reflecting moderate engagement levels. The remaining topics are in group ``d'', which exhibits the lowest engagement levels.

\begin{figure}[t]
	\centering
	\includegraphics[width=\columnwidth]{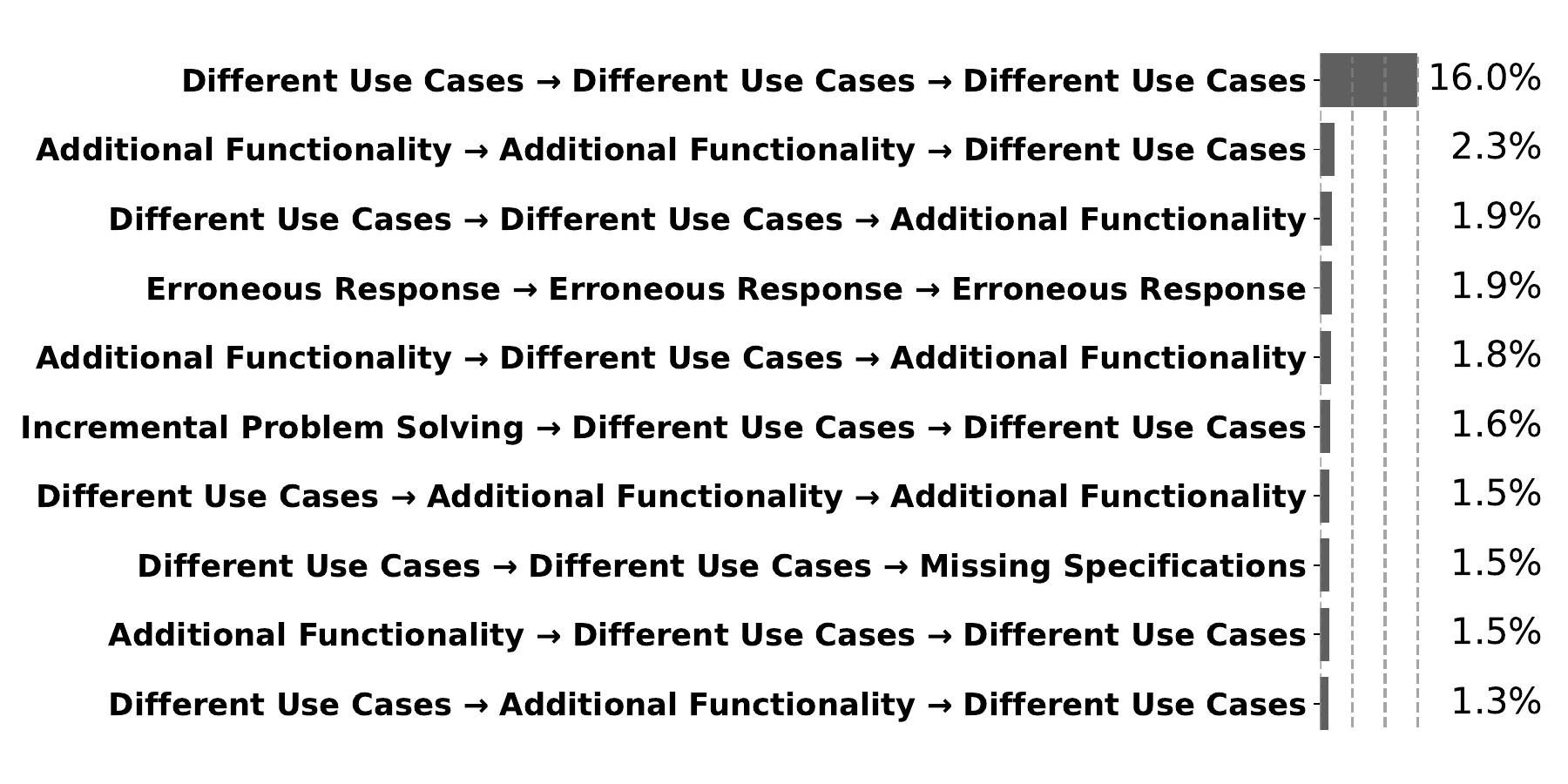}
    \vspace{-25pt}
	\caption{Most frequent 3-gram ``chain-of-gap'' patterns in developer-LLM conversations within ``Topic 8'', the cluster with the highest average turn count. Each 3-gram represents a sequence of consecutive prompt design gaps.}
	\label{fig:RQ2_WhyLongTurn}
    \vspace{-10pt}
\end{figure}

\textbf{In conversations on the topic with the highest TC~(Topic~8), the most frequent prompt gap sequence involves three successive occurrences of Different Use Cases, accounting for 16.0\% of the observed 3-gram chains.} As shown in Figure~\ref{fig:RQ2_WhyLongTurn}, this pattern involves repeated shifts in user intent or use cases and is the most frequent three-gap pattern within Topic 8, the topic with the highest mean turn count. The next most common sequences are Additional Functionality $\rightarrow$ Additional Functionality $\rightarrow$ Different Use Cases (2.3\%) and Different Use Cases $\rightarrow$ Different Use Cases $\rightarrow$ Additional Functionality (1.9\%), both suggesting alternation between adding features and switching use cases before objectives are completed. We also observe Erroneous Response $\rightarrow$ Erroneous Response $\rightarrow$ Erroneous Response (1.9\%), indicating consecutive incorrect LLM outputs that are not effectively corrected and therefore extend the conversation.

\smallskip
\begin{Summary}{Summary of RQ1}{}{Developers frequently seek LLM assistance for \textit{Web design and development}~(9.6\%), \textit{Machine learning model training and AI bot deployment}~(8.7\%), and \textit{Assembly-level operations and memory handling}~(4.9\%). Moreover, repeated shifts in use cases (16.0\%) and alternation between features and objectives are prevalent prompt gaps in longer conversations.}
\end{Summary}

\subsection{\textbf{RQ2: \rqthree}}\label{sec:RQ3}

\motivation As developers increasingly rely on LLMs to solve code-related tasks, the quality of LLM-generated code becomes a critical concern.  While LLMs have demonstrated impressive capabilities, the generated code may contain defects or fail to follow best practices. A comprehensive evaluation of LLM-generated code across different programming languages and real-world scenarios can provide valuable insights into their strengths and weaknesses. The findings of this RQ can help developers make more informed and responsible usage of LLMs, guide code assistant developers to address identified issues, and offer research directions for improving the capabilities and reliability of LLMs in code-related tasks.

\begin{figure*}[ht]
    \centering
    \includegraphics[width=\textwidth]{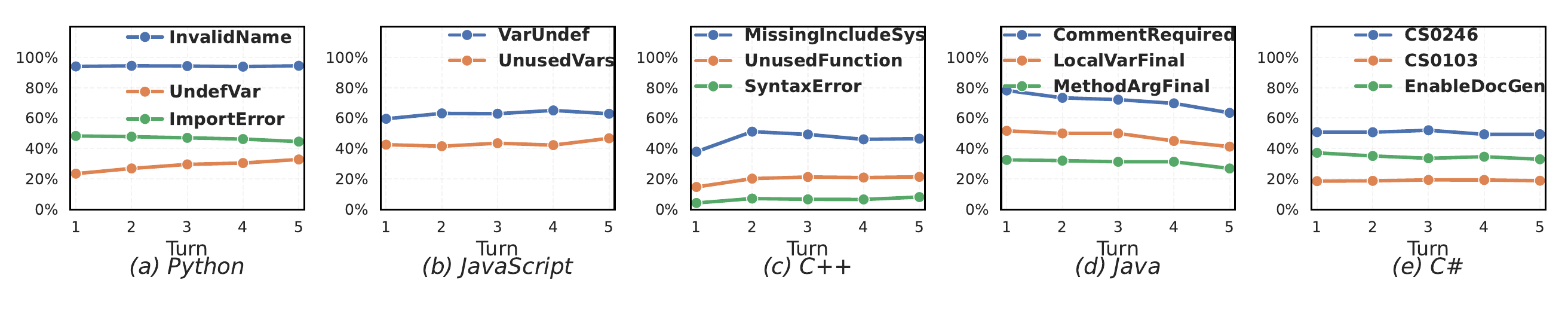}
    \vspace{-25pt}
    \caption{Distribution of the top three quality issues across turns in only multi-turn conversations, 
    ranked by their frequency at the first turn.}
    \label{fig:RQ3_3SingleIssueCompare}
    \vspace{-10pt}
\end{figure*}

\findings \textbf{In Python, invalid naming is the most frequent quality issue, affecting 83.4\% of LLM-generated snippets; across multi-turn conversations, the proportion of undefined-variable issues increases, while the proportion of import-error issues decreases.} 
As shown in Table~\ref{tab:code-linter-percentage}, 83.4\% of the first-turn Python snippets contain naming issues that fail to comply with Pylint's default PEP8 rules~\cite{pep8}. Undefined variables~(E0602) occur in 30.8\% of snippets and unused imports~(W0611) in 9.4\% of snippets. In human-written Python code from Jupyter notebooks~\cite{10356405}, import errors account for 61.2\% of error-type issues, undefined variables for 18.9\%, and unused imports for 24.1\% of warning-type issues. The same issue types are prominent in our LLM-generated snippets, indicating that they are not unique to LLMs but are general code-quality problems. In follow-up multi-turn Python conversations (Figure~\ref{fig:RQ3_3SingleIssueCompare}), the prevalence of \textit{UndefinedVar} increases from 23.5\% of code snippets at turn 1 to 32.8\% at turn 5 ($p<0.05$), indicating challenges in preserving variable context, while the proportion of \textit{ImportError} decreases from 48.3\% of code snippets to 44.6\% ($p<0.05$), suggesting missing imports are progressively added.

\begin{table}[t]
\centering
\caption{Static analysis results of the first-turn LLM-generated code snippets across all conversations (single- and multi-turn).}
\label{tab:code-linter-percentage}
\setlength{\tabcolsep}{0.5\tabcolsep}
\adjustbox{max width=\columnwidth}{
\begin{tabular}{@{}cllrll@{}}
\toprule
                         & \textbf{Severity}            & \textbf{Top 3 Quality Issues}                & \textbf{\begin{tabular}[c]{@{}r@{}}\%Code\\Snippet\end{tabular}} & \textbf{\begin{tabular}[c]{@{}l@{}}\% in\\Human-Code*\end{tabular}} & \textbf{Study} \\ \midrule
\multirow{11}{*}{\rotatebox[origin=c]{90}{\textbf{Python}}}  & \multirow{3}{*}{Error}       & E0602:UndefinedVariable           & 30.8\%   & 18.9\% of errors      & \cite{10356405}               \\
                         &                              & E0401:ImportError                 & 20.8\%   & 61.2\% of errors      & \cite{10356405}               \\
                         &                              & E0001:SyntaxError                 & 8.0\%    & \textit{n/r}          &                               \\ \cmidrule(l){2-6}
                         & \multirow{3}{*}{Warning}     & W0621:RedefinedOuterName          & 9.9\%    & 21.1\% of warnings    & \cite{10356405}               \\
                         &                              & W0611:UnusedImport                & 9.4\%    & 24.1\% of warnings    & \cite{10356405}               \\
                         &                              & W0612:UnusedVariable              & 3.7\%    & \textit{n/r}          &                               \\ \cmidrule(l){2-6}
                         & \multirow{2}{*}{Refactor}    & R0903:TooFewPublicMethods         & 4.2\%    & \textit{n/r}          &                               \\
                         &                              & R1705:NoElseReturn                & 2.6\%    & 13.2\% of refactors   & \cite{10356405}               \\ \cmidrule(l){2-6}
                         & \multirow{3}{*}{Convention}  & C0103:InvalidName                 & 83.4\%   & 27.7\% of convention & \cite{10356405}               \\
                         &                              & C0116:MissingFunctionDocs         & 32.1\%   & 5.9\% of convention  & \cite{10356405}               \\
                         &                              & C0301:LineTooLong                 & 15.9\%   & 18.2\% of convention & \cite{10356405}               \\ \midrule
\multirow{3}{*}{\rotatebox[origin=c]{90}{\textbf{JS}}}      & \multirow{2}{*}{Error}       & NoUndef                           & 75.3\%   & 91.0\% of variables   & \cite{8816735}                \\
                         &                              & SyntaxError                       & 14.4\%   & \textit{n/r}          &                               \\ \cmidrule(l){2-6}
                         & Warning                      & NoUnusedVars                      & 33.7\%   & 8.6\% of variables    & \cite{8816735}                \\ \midrule
\multirow{4}{*}{\rotatebox[origin=c]{90}{\textbf{C++}}}     & Error                        & SyntaxError                       & 9.4\%    & 20.0\% of snippets    & \cite{9359361}                \\ \cmidrule(l){2-6}
                         & Style                        & UnusedFunction                    & 17.8\%   & \textit{n/r}          &                               \\ \cmidrule(l){2-6}
                         & \multirow{2}{*}{Info.}       & MissingIncludeSystem              & 41.1\%   & \textit{n/r}          &                               \\
                         &                              & MissingInclude                    & 6.6\%    & \textit{n/r}          &                               \\ \midrule
\multirow{6}{*}{\rotatebox[origin=c]{90}{\textbf{Java}}}   & High                         & BestPractices:SystemPrintln       & 22.4\%   & \textit{n/r}          &                               \\ \cmidrule(l){2-6}
                         & \multirow{3}{*}{Medium}      & Documents:CommentRequired         & 75.9\%   & 42.2\% of PRs         & \cite{LENARDUZZI2021110806}   \\
                         &                              & Design:UseUtilityClass            & 47.5\%   & \textit{n/r}          &                               \\
                         &                              & Style:LocalVarCouldBeFinal        & 45.3\%   & 41.1\% of PRs         & \cite{LENARDUZZI2021110806}   \\ \cmidrule(l){2-6}
                         & \multirow{2}{*}{Low}         & Style:ShortClassName              & 8.2\%    & \textit{n/r}          &                               \\
                         &                              & Style:UnnecessaryImport           & 5.3\%    & \textit{n/r}          &                               \\
\midrule
\multirow{9}{*}{\rotatebox[origin=c]{90}{\textbf{C\#}}}     & \multirow{3}{*}{Error}       & CS0246:NamespaceUnfound           & 49.2\%   & \textit{n/r}*         & \cite{csharp_humancode_so}    \\
                         &                              & CS0103:NameIdentifierNotExist     & 24.3\%   & \textit{n/r}*         & \cite{csharp_humancode_so}    \\
                         &                              & CS0106:ModifierNotValid           & 17.7\%   & \textit{n/r}*         & \cite{csharp_humancode_so}    \\ \cmidrule(l){2-6}
                         & \multirow{3}{*}{Warning}     & EnableGenDocFile                  & 42.4\%   & \textit{n/r}          &                               \\
                         &                              & CA1852:SealInternalTypes          & 9.6\%    & \textit{n/r}          &                               \\
                         &                              & CA1050:TypesInNamespaces          & 5.2\%    & \textit{n/r}          &                               \\ \cmidrule(l){2-6}
                         & \multirow{3}{*}{Style}       & IDE0040:AddAccessModifiers        & 10.0\%   & \textit{n/r}          &                               \\
                         &                              & IDE0055:FormattingRule            & 9.8\%    & \textit{n/r}          &                               \\
                         &                              & IDE0210:ToTopLevelStatements      & 8.1\%    & \textit{n/r}          &                               \\ \bottomrule
\end{tabular}
}
\vspace{2pt}\par\noindent\begin{minipage}{\columnwidth}
\raggedright\scriptsize \textbf{Human-Code}: human-written code quality from prior work; each value states its own base (\% of a severity category, \% of variable violations, \% of snippets, or \% of pull requests). \textit{n/r} = not reported by available prior work. For C\# (\textit{n/r}*), the cited study~\cite{csharp_humancode_so} reports only aggregate compilability, so no per-issue value exists. Info.: Information.
\end{minipage}
\vspace{-15pt}
\end{table}

\textbf{In JavaScript, undefined variables are the most frequent quality issue, affecting 75.3\% of LLM-generated snippets; unused variables occur in 33.7\%, and syntax errors in 14.4\%, with no significant changes across turns.} 
As shown in Table~\ref{tab:code-linter-percentage}, undefined variables appear in 75.3\% of first-turn JavaScript snippets. In human-written JavaScript on Stack Overflow, undefined variables are the dominant variable-related issue, accounting for 91.0\% of variable-category violations~\cite{8816735}. Syntax errors are found in 14.4\% of snippets, while the study on human-written code~\cite{8816735} analyzes only parseable code and reports no syntax-error rate, so no human comparison is available. In LLM-generated JavaScript, the most frequent cause of these syntax errors is unexpected angle brackets, which occur in 5.8\% of code snippets. In multi-turn JavaScript conversations (Figure~\ref{fig:RQ3_3SingleIssueCompare}), the prevalence of top issues remains statistically unchanged.

\textbf{In C++, missing \texttt{\#include} headers are the most frequent quality issue, affecting 41.1\% of initial-turn LLM-generated snippets; unused functions occur in 17.8\%, and syntax errors in 9.4\%, with no significant changes across turns.} As shown in Table~\ref{tab:code-linter-percentage}, syntax errors are present in 9.4\% of snippets, which is less than the 20.0\% reported for human-written C/C++ code on Stack Overflow when analyzed with CppCheck~\cite{9359361}. In follow-up multi-turn C++ conversations (Figure~\ref{fig:RQ3_3SingleIssueCompare}), the prevalence of \textit{MissingIncludeSystem} rises from 37.9\% of code snippets at turn 1 to 46.5\% at turn 5, and \textit{SyntaxError} from 4.2\% to 8.1\%, but neither change is statistically significant.

\textbf{In Java, missing required comments are the most frequent quality issue, affecting 75.9\% of initial-turn LLM-generated snippets, with documentation violations and missed opportunities to declare local variables as \texttt{final} both declining across turns.}
As shown in Table~\ref{tab:code-linter-percentage}, absence of \texttt{final} on local variables (\textit{LocalVariableCouldBeFinal}) occurs in 45.3\% of snippets. The same issue is common in human-written Java, appearing in 41.1\% of pull requests across 28 large projects~\cite{LENARDUZZI2021110806}. In follow-up multi-turn Java conversations (Figure~\ref{fig:RQ3_3SingleIssueCompare}), the prevalence of \textit{CommentRequired} decreases from 78.1\% of code snippets at turn 1 to 63.4\% at turn 5 ($p<0.05$), and \textit{LocalVariableCouldBeFinal} from 51.6\% to 41.3\% ($p<0.05$), suggesting iterative improvements in documentation and in declaring local variables as \texttt{final}.

\textbf{Finally, in C\#, unresolved namespaces are the most frequent quality issue, affecting 49.2\% of initial-turn LLM-generated snippets. Across multi-turn conversations, their prevalence remains unchanged~($p>0.05$).} As shown in Table~\ref{tab:code-linter-percentage}, missing documentation occurs in 42.4\% of code snippets, which can hinder code comprehension and maintenance. Other notable issues include omitted accessibility modifiers~(10.0\%) and formatting violations (9.8\%), both of which impact code readability. In human-written C\# code from Stack Overflow~\cite{csharp_humancode_so}, only 16.0\% of snippets are parsable and 0.12\% are compilable, reflecting overall usability rather than the per-issue rates reported here.

\begin{table}[t]
\centering
\caption{Categories of follow-up prompts preceding syntax-error-free responses, adapted from \cite{11025620}}
\label{tab:syntax_error_reduction_categories}
\begin{adjustbox}{width=\columnwidth}
\begin{tabular}{@{}lrr@{}}
\toprule
\textbf{Category Name} & \textbf{Frequency} & \textbf{Percentage} \\ 
\midrule
Point out mistake then request fix       & 88 & 23.0\% \\
Ask questions to guide correct solution  & 65 & 17.0\% \\
Add specific instructions                & 63 & 16.4\% \\
Request improvements                     & 53 & 13.8\% \\
\textbf{Request Full Code$^{*}$}         & \textbf{42} & \textbf{11.0\%} \\
Add more context                         & 27 & 7.0\% \\
Request more description about code      & 17 & 4.4\% \\
\textbf{Change Use Cases$^{*}$}          & \textbf{12} & \textbf{3.1\%} \\
Request examples                         & 8  & 2.1\% \\
Request alternative code generation      & 4  & 1.0\% \\
Request verification                     & 4  & 1.0\% \\
\midrule
\textbf{Total}                           & \textbf{383} & \textbf{100\%} \\
\bottomrule
\multicolumn{3}{@{}l}{\scriptsize $^{*}$ Categories defined in this study.}
\end{tabular}
\end{adjustbox}
\vspace{-15pt}
\end{table}

\textbf{``Point out mistake then request fix'' is the most frequent follow-up prompt type, occurring in 23.0\% of the 383 sampled transitions.} As listed in Table~\ref{tab:syntax_error_reduction_categories}, explicitly naming the error and requesting a fix frequently precedes syntactically valid code generation. The next most frequent prompt types are \textit{Ask questions to guide correct solution} (17.0\%) and \textit{Add specific instructions} (16.4\%), both of which clarify the developer's intent or specify concrete edits. Together, the three prompt types account for 56.4\% of the resolved cases, suggesting that clear error signaling, guided questioning, and precise directives are prevalent in successful resolutions and may reduce turn count when seeking non-syntax-error outputs.

\smallskip
\begin{Summary}{Summary of RQ2}{}{The most frequent issues in LLM-generated code are: invalid naming in Python, undefined variables in JavaScript, missing \#include in C++, missing required comments in Java, and namespace unfound errors in C\# (49.2\%). Moreover, in multi-turn conversations, Python import errors decrease from turn 1 to turn 5, while Java documentation violations drop from 78.1\% to 63.4\%.}
\end{Summary}

\section{Discussion}\label{sec:implications}

In this section, we discuss the implications of our study for practitioners, tool builders, and researchers in software engineering.


\textbf{Proactively managing evolving developer intent in multi-turn coding conversations.} Our study finds that prolonged conversations are often driven by recurring intent shifts, with repeated changes in use cases forming the most common gap pattern. Although current assistants maintain context and follow the latest instruction, they react to each prompt in isolation rather than tracking how a developer's goal evolves, leaving redirected intent unmanaged and prolonging interactions. We therefore recommend tool builders to equip code assistants with mechanisms that detect intent shifts and confirm the redirected goal before continuing, and researchers to model intent evolution across turns.

\textbf{Reusing real-world multi-turn coding interactions to fine-tune and evaluate code assistants.} \dataset captures real developer-LLM coding sessions across more than 20 programming languages and multiple turns, including conversations that extend to ten turns or more. Unlike curated single-turn benchmarks, it reflects how developers phrase, refine, and repair requests in practice. We therefore encourage researchers and tool builders to use \dataset to fine-tune smaller, task-specialized code models and to evaluate assistants under realistic multi-turn and long-context conditions. Future studies could also extend our static analysis to security and vulnerability assessment, or compare code generated by other and newer models against our ChatGPT-based baseline.

\textbf{Using real-world conversations to build benchmarks for multi-turn coding assistants.} \dataset captures how developers formulate, refine, redirect, and repair coding requests across multiple turns and more than 20 programming languages. Researchers can use these interactions to construct evaluation benchmarks that measure conversation-level behavior, including intent-shift recovery, code-quality regressions, error-correction rates, turns to resolution, and robustness to long contexts. These measures would complement existing single-turn benchmarks by evaluating whether an assistant preserves requirements and improves its outputs throughout a conversation. The dataset may also support fine-tuning task-specific models, but its primary value is enabling realistic and reproducible evaluation of multi-turn coding assistants.

\section{Threats to Validity}\label{sec:threadstovalidity}


~\textbf{Internal Validity.} We select static analysis tools that are widely recognized for their reliability and relevance to the programming languages under investigation. However, a potential internal validity threat arises from the specific configuration settings used for each static analysis tool. To mitigate this concern, we follow the official documentation and recommended guidelines for each linter, thereby ensuring consistency, reliability, and transparency throughout our evaluation process.

Another limitation is that our code quality assessment is restricted to the types of issues supported by static analyzers. 
Other quality attributes, such as code complexity and security vulnerabilities, are beyond the scope of our current analysis and are therefore not reflected in our findings.


~\textbf{External Validity.} \dataset is derived from WildChat~\cite{zhao2024wildchat}, whose chatbot is hosted on Hugging Face. Its users are therefore likely to skew toward the IT community and specific communities, which may over-represent coding-related questions and particular interaction styles. As a result, the topic distribution characterizes this population rather than developers in general. In addition, \dataset captures conversations with ChatGPT models~(from GPT-3.5 to o1) collected between April 2023 and July 2025. Consequently, the reported results reflect these models in these period. The analysis framework may be applied to conversations from other LLMs, such as Claude and Gemini, but the specific frequencies and trends may differ because of changes in model capabilities. 

Finally, our findings may not generalize to repository-based coding agents, such as Codex and Claude Code, which can access project files, execute code, and revise outputs using tool feedback. \dataset instead captures prompt-response interactions between users engaged in coding conversations and conversational coding assistants.

\section{Related Work}\label{sec:relatedwork}

\textbf{Developer-LLM Interaction Datasets.} Prior studies rely on curated or self-reported sources. Several evaluate LLMs on competitive-programming problems from LeetCode~\cite{10.1145/3661167.3661221,10621717} or on questions mined from Stack Overflow~\cite{DBLP:journals/corr/abs-2308-02312,10260753,dasilva2024chatgptvsllamaimpact}. Others analyze developer-ChatGPT interactions collected from GitHub, categorizing prompt types~\cite{hao2024empirical} or discussion topics~\cite{10.1145/3643991.3645080,das2024developersengagechatgptissuetracker}, while DevGPT~\cite{xiao2024devgpt} provides 29,778 conversations and 19,106 code snippets from GitHub and Hacker News. Unlike these curated or platform-centric sources, \dataset (Table~\ref{tab:rw_dataset}) comprises 587,568 unconstrained, unedited developer-LLM conversations from a public chatbot service, built on WildChat~\cite{zhao2024wildchat}, and thus reflects a broader range of everyday tasks than the specialized prompts shared on GitHub.

\textbf{Code Quality of LLM-Generated Code.} Recent studies assess the quality of LLM-generated code. \cite{liu2024refining} examine issues in ChatGPT solutions for 2,033 LeetCode tasks, \cite{siddiq2024quality} apply static analysis to Python and Java snippets from DevGPT, and \cite{10870152} evaluate documentation and unused variables across 500 tasks. In contrast, we analyze code across five languages drawn from real-world conversations and, beyond single-turn settings, track how quality issues evolve across consecutive turns as developers iteratively interact with LLMs.

\section{Conclusion}\label{sec:conclusion}

In this study, we introduce \dataset, a filtered subset of the WildChat dataset containing 587,568 real-world developer-LLM conversations with 1.7 million code snippets across more than 20 programming languages. Analyzing these conversations, we find that developers most often seek help with tasks such as web design, machine learning model training, and low-level memory operations, and that prolonged conversations are frequently driven by shifting use cases and iterative refinement. Examining the generated code, we find that it frequently exhibits quality issues, including syntax errors, undefined variables, and maintainability problems, and that some of these issues evolve across successive turns, such as undefined variables, becoming more prevalent as conversations progress. These findings suggest that conversational assistants would benefit from tracking evolving developer intent and monitoring code quality across turns. In future work, we plan to develop domain-specific benchmarks for the most frequently requested tasks to improve the usability and trustworthiness of LLM-generated code.

\balance
\bibliographystyle{IEEEtranBST2/IEEEtran}
\bibliography{Bib}

@inproceedings{zhao2024wildchat,
  author       = {Wenting Zhao and
                  Xiang Ren and
                  Jack Hessel and
                  Claire Cardie and
                  Yejin Choi and
                  Yuntian Deng},
  title        = {WildChat: 1M ChatGPT Interaction Logs in the Wild},
  booktitle    = {The Twelfth International Conference on Learning Representations,
                  {ICLR} 2024, Vienna, Austria, May 7-11, 2024},
  publisher    = {OpenReview.net},
  year         = {2024},
  url          = {https://openreview.net/forum?id=Bl8u7ZRlbM},
  bibsource    = {dblp computer science bibliography, https://dblp.org}
}

@inproceedings{zheng2023lmsys,
	author       = {Lianmin Zheng and
                  Wei{-}Lin Chiang and
                  Ying Sheng and
                  Tianle Li and
                  Siyuan Zhuang and
                  Zhanghao Wu and
                  Yonghao Zhuang and
                  Zhuohan Li and
                  Zi Lin and
                  Eric P. Xing and
                  Joseph E. Gonzalez and
                  Ion Stoica and
                  Hao Zhang},
  title        = {LMSYS-Chat-1M: {A} Large-Scale Real-World {LLM} Conversation Dataset},
  booktitle    = {The Twelfth International Conference on Learning Representations,
                  {ICLR} 2024, Vienna, Austria, May 7-11, 2024},
  publisher    = {OpenReview.net},
  year         = {2024},
  url          = {https://openreview.net/forum?id=BOfDKxfwt0},
  bibsource    = {dblp computer science bibliography, https://dblp.org}
}

@inproceedings{kopf2024openassistant,
	 author = {K\"{o}pf, Andreas and Kilcher, Yannic and von R\"{u}tte, Dimitri and Anagnostidis, Sotiris and Tam, Zhi Rui and Stevens, Keith and Barhoum, Abdullah and Nguyen, Duc and Stanley, Oliver and Nagyfi, Rich\'{a}rd and ES, Shahul and Suri, Sameer and Glushkov, David and Dantuluri, Arnav and Maguire, Andrew and Schuhmann, Christoph and Nguyen, Huu and Mattick, Alexander},
 booktitle = {Advances in Neural Information Processing Systems},
 editor = {A. Oh and T. Naumann and A. Globerson and K. Saenko and M. Hardt and S. Levine},
 pages = {47669--47681},
 publisher = {Curran Associates, Inc.},
 title = {OpenAssistant Conversations - Democratizing Large Language Model Alignment},
 url = {http://papers.nips.cc/paper\_files/paper/2023/hash/949f0f8f32267d297c2d4e3ee10a2e7e-Abstract-Datasets\_and\_Benchmarks.html},
 volume = {36},
 year = {2023}
}

@article{liu2024refining,
	author = {Liu, Yue and Le-Cong, Thanh and Widyasari, Ratnadira and Tantithamthavorn, Chakkrit and Li, Li and Le, Xuan-Bach D. and Lo, David},
title = {Refining ChatGPT-Generated Code: Characterizing and Mitigating Code Quality Issues},
year = {2024},
issue_date = {June 2024},
publisher = {Association for Computing Machinery},
address = {New York, NY, USA},
volume = {33},
number = {5},
issn = {1049-331X},
url = {https://doi.org/10.1145/3643674},
doi = {10.1145/3643674},
month = jun,
articleno = {116},
numpages = {26}
}

@inproceedings{siddiq2024quality,
	author = {Siddiq, Mohammed Latif and Roney, Lindsay and Zhang, Jiahao and Santos, Joanna Cecilia Da Silva},
title = {Quality Assessment of ChatGPT Generated Code and their Use by Developers},
year = {2024},
isbn = {9798400705878},
publisher = {Association for Computing Machinery},
address = {New York, NY, USA},
url = {https://doi.org/10.1145/3643991.3645071},
doi = {10.1145/3643991.3645071},
booktitle = {Proceedings of the 21st International Conference on Mining Software Repositories},
pages = {152–156},
numpages = {5},
location = {Lisbon, Portugal},
series = {MSR '24}
}

@inproceedings{xiao2024devgpt,
	author = {Xiao, Tao and Treude, Christoph and Hata, Hideaki and Matsumoto, Kenichi},
title = {DevGPT: Studying Developer-ChatGPT Conversations},
year = {2024},
isbn = {9798400705878},
publisher = {Association for Computing Machinery},
address = {New York, NY, USA},
url = {https://doi.org/10.1145/3643991.3648400},
doi = {10.1145/3643991.3648400},
booktitle = {Proceedings of the 21st International Conference on Mining Software Repositories},
pages = {227–230},
numpages = {4},
location = {Lisbon, Portugal},
series = {MSR '24}
}

@article{hao2024empirical,
	author = {Hao, Huizi and Hasan, Kazi Amit and Qin, Hong and Macedo, Marcos and Tian, Yuan and Ding, Steven H. H. and Hassan, Ahmed E.},
title = {An empirical study on developers’ shared conversations with ChatGPT in GitHub pull requests and issues},
year = {2024},
issue_date = {Nov 2024},
publisher = {Kluwer Academic Publishers},
address = {USA},
volume = {29},
number = {6},
issn = {1382-3256},
url = {https://doi.org/10.1007/s10664-024-10540-x},
doi = {10.1007/s10664-024-10540-x},
journal = {Empirical Softw. Engg.},
month = sep,
numpages = {43}
}

@article{jiang2024surveylargelanguagemodels,
author = {Jiang, Juyong and Wang, Fan and Shen, Jiasi and Kim, Sungju and Kim, Sunghun},
title = {A Survey on Large Language Models for Code Generation},
year = {2026},
issue_date = {February 2026},
publisher = {Association for Computing Machinery},
address = {New York, NY, USA},
volume = {35},
number = {2},
issn = {1049-331X},
url = {https://doi.org/10.1145/3747588},
doi = {10.1145/3747588},
journal = {ACM Trans. Softw. Eng. Methodol.},
month = jan,
articleno = {58},
numpages = {72}
}

@inproceedings{10.1145/3701625.3701650,
	author = {Sim\~{o}es, Igor Regis da Silva and Venson, Elaine},
title = {Evaluating Source Code Quality with Large Language Models: a comparative study},
year = {2024},
isbn = {9798400717772},
publisher = {Association for Computing Machinery},
address = {New York, NY, USA},
url = {https://doi.org/10.1145/3701625.3701650},
doi = {10.1145/3701625.3701650},
booktitle = {Proceedings of the XXIII Brazilian Symposium on Software Quality},
pages = {103–113},
numpages = {11},
location = {
},
series = {SBQS '24}
}

@article{haider2024promptingfinetuninglargelanguage,
	author       = {Md. Asif Haider and
                  Ayesha Binte Mostofa and
                  Sk Md Mosaddek Hossain and
                  Anindya Iqbal and
                  Toufique Ahmed},
  title        = {Prompting and Fine-tuning Large Language Models for Automated Code
                  Review Comment Generation},
  journal      = {CoRR},
  volume       = {abs/2411.10129},
  year         = {2024},
  url          = {https://doi.org/10.48550/arXiv.2411.10129},
  doi          = {10.48550/ARXIV.2411.10129},
  eprinttype   = {arXiv},
  eprint       = {2411.10129},
  bibsource    = {dblp computer science bibliography, https://dblp.org}
}

@article{10.1145/3697012,
author = {Gu, Xiaodong and Chen, Meng and Lin, Yalan and Hu, Yuhan and Zhang, Hongyu and Wan, Chengcheng and Wei, Zhao and Xu, Yong and Wang, Juhong},
title = {On the Effectiveness of Large Language Models in Domain-Specific Code Generation},
year = {2025},
issue_date = {March 2025},
publisher = {Association for Computing Machinery},
address = {New York, NY, USA},
volume = {34},
number = {3},
issn = {1049-331X},
url = {https://doi.org/10.1145/3697012},
doi = {10.1145/3697012},
journal = {ACM Trans. Softw. Eng. Methodol.},
month = feb,
articleno = {78},
numpages = {22}
}

@misc{vicuna2023,
	title        = {{Vicuna: An Open-Source Chatbot Impressing GPT-4 with 90\% ChatGPT Quality}},
	author       = {The Vicuna Team},
	year         = 2023,
	month        = {March},
    url={https://www.lmsys.org/blog/2023-03-30-vicuna/}
}

@inproceedings{10196869,
	author={Feng, Yunhe and Vanam, Sreecharan and Cherukupally, Manasa and Zheng, Weijian and Qiu, Meikang and Chen, Haihua},
  booktitle={2023 IEEE 47th Annual Computers, Software, and Applications Conference (COMPSAC)}, 
  title={Investigating Code Generation Performance of ChatGPT with Crowdsourcing Social Data}, 
  year={2023},
  volume={},
  number={},
  pages={876-885},
  doi={10.1109/COMPSAC57700.2023.00117},
  url={https://doi.org/10.1109/COMPSAC57700.2023.00117}}

@article{10870152,
	author={Almanasra, Sally and Suwais, Khaled},
  journal={IEEE Access}, 
  title={Analysis of ChatGPT-Generated Codes Across Multiple Programming Languages}, 
  year={2025},
  volume={13},
  number={},
  pages={23580-23596},
  doi={10.1109/ACCESS.2025.3538050},
  url={https://doi.org/10.1109/ACCESS.2025.3538050}}

@inproceedings{10.1145/3643991.3645080,
	author = {Sagdic, Ertugrul and Bayram, Arda and Islam, Md Rakibul},
title = {On the Taxonomy of Developers' Discussion Topics with ChatGPT},
year = {2024},
isbn = {9798400705878},
publisher = {Association for Computing Machinery},
address = {New York, NY, USA},
url = {https://doi.org/10.1145/3643991.3645080},
doi = {10.1145/3643991.3645080},
booktitle = {Proceedings of the 21st International Conference on Mining Software Repositories},
pages = {197–201},
numpages = {5},
location = {Lisbon, Portugal},
series = {MSR '24}
}

@inproceedings{DBLP:journals/corr/abs-2308-02312,
	author = {Kabir, Samia and Udo-Imeh, David N. and Kou, Bonan and Zhang, Tianyi},
title = {Is Stack Overflow Obsolete? An Empirical Study of the Characteristics of ChatGPT Answers to Stack Overflow Questions},
year = {2024},
isbn = {9798400703300},
publisher = {Association for Computing Machinery},
address = {New York, NY, USA},
url = {https://doi.org/10.1145/3613904.3642596},
doi = {10.1145/3613904.3642596},
articleno = {935},
numpages = {17},
location = {Honolulu, HI, USA},
series = {CHI '24}
}

@inproceedings{10.1145/3661167.3661221,
	author       = {Tristan Coignion and
                  Cl{\'{e}}ment Quinton and
                  Romain Rouvoy},
  title        = {A Performance Study of LLM-Generated Code on Leetcode},
  booktitle    = {Proceedings of the 28th International Conference on Evaluation and
                  Assessment in Software Engineering, {EASE} 2024, Salerno, Italy, June
                  18-21, 2024},
  pages        = {79--89},
  publisher    = {{ACM}},
  year         = {2024},
  url          = {https://doi.org/10.1145/3661167.3661221},
  doi          = {10.1145/3661167.3661221},
  bibsource    = {dblp computer science bibliography, https://dblp.org}
}

@inproceedings{10260753,
	author={Delile, Zack and Radel, Sean and Godinez, Joe and Engstrom, Garrett and Brucker, Theo and Young, Kenzie and Ghanavati, Sepideh},
  booktitle={2023 IEEE 31st International Requirements Engineering Conference Workshops (REW)}, 
  title={Evaluating Privacy Questions from Stack Overflow: Can ChatGPT Compete?}, 
  year={2023},
  volume={},
  number={},
  pages={239-244},
  doi={10.1109/REW57809.2023.00048},
  url={https://doi.org/10.1109/REW57809.2023.00048}}

@misc{dasilva2024chatgptvsllamaimpact,
	  author       = {L{\'{e}}uson M. P. da Silva and
                  Jordan Samhi and
                  Foutse Khomh},
  title        = {ChatGPT vs LLaMA: Impact, Reliability, and Challenges in Stack Overflow
                  Discussions},
  journal      = {CoRR},
  volume       = {abs/2402.08801},
  year         = {2024},
  url          = {https://doi.org/10.48550/arXiv.2402.08801},
  doi          = {10.48550/ARXIV.2402.08801},
  eprinttype   = {arXiv},
  eprint       = {2402.08801},
  bibsource    = {dblp computer science bibliography, https://dblp.org}
}

@inproceedings{10621717,
	author={Nikolaidis, Nikolaos and Flamos, Karolos and Gulati, Khanak and Feitosa, Daniel and Ampatzoglou, Apostolos and Chatzigeorgiou, Alexander},
  booktitle={2024 IEEE International Conference on Software Analysis, Evolution and Reengineering - Companion (SANER-C)}, 
  title={A Comparison of the Effectiveness of ChatGPT and Co-Pilot for Generating Quality Python Code Solutions}, 
  year={2024},
  volume={},
  number={},
  pages={93-101},
  doi={10.1109/SANER-C62648.2024.00018},
url={https://doi.org/10.1109/SANER-C62648.2024.00018}
}

@INPROCEEDINGS{das2024developersengagechatgptissuetracker,
	author={Das, Joy Krishan and Mondal, Saikat and Roy, Chanchal K.},
  booktitle={2025 IEEE International Conference on Software Analysis, Evolution and Reengineering (SANER)}, 
  title={Why Do Developers Engage with ChatGPT in Issue-Tracker? Investigating Usage and Reliance on ChatGPT-Generated Code}, 
  year={2025},
  volume={},
  number={},
  pages={68-79},
  doi={10.1109/SANER64311.2025.00015},
  url={https://doi.org/10.1109/SANER64311.2025.00015}
  }

@inproceedings{10.1145/2684822.2685324,
	author = {R\"{o}der, Michael and Both, Andreas and Hinneburg, Alexander},
title = {Exploring the Space of Topic Coherence Measures},
year = {2015},
isbn = {9781450333177},
publisher = {Association for Computing Machinery},
address = {New York, NY, USA},
url = {https://doi.org/10.1145/2684822.2685324},
doi = {10.1145/2684822.2685324},
booktitle = {Proceedings of the Eighth ACM International Conference on Web Search and Data Mining},
pages = {399–408},
numpages = {10},
location = {Shanghai, China},
series = {WSDM '15}
}

@inproceedings{NIPS2017_3f5ee243,
  author       = {Ashish Vaswani and
                  Noam Shazeer and
                  Niki Parmar and
                  Jakob Uszkoreit and
                  Llion Jones and
                  Aidan N. Gomez and
                  Lukasz Kaiser and
                  Illia Polosukhin},
  editor       = {Isabelle Guyon and
                  Ulrike von Luxburg and
                  Samy Bengio and
                  Hanna M. Wallach and
                  Rob Fergus and
                  S. V. N. Vishwanathan and
                  Roman Garnett},
  title        = {Attention is All you Need},
  booktitle    = {Advances in Neural Information Processing Systems 30: Annual Conference
                  on Neural Information Processing Systems 2017, December 4-9, 2017,
                  Long Beach, CA, {USA}},
  pages        = {5998--6008},
  year         = {2017},
  url          = {https://proceedings.neurips.cc/paper/2017/hash/3f5ee243547dee91fbd053c1c4a845aa-Abstract.html},
  bibsource    = {dblp computer science bibliography, https://dblp.org}
}

@inproceedings{10.1145/3643991.3645085,
 author = {Mondal, Saikat and Bappon, Suborno Deb and Roy, Chanchal K.},
title = {Enhancing User Interaction in ChatGPT: Characterizing and Consolidating Multiple Prompts for Issue Resolution},
year = {2024},
isbn = {9798400705878},
publisher = {Association for Computing Machinery},
address = {New York, NY, USA},
url = {https://doi.org/10.1145/3643991.3645085},
doi = {10.1145/3643991.3645085},
booktitle = {Proceedings of the 21st International Conference on Mining Software Repositories},
pages = {222–226},
numpages = {5},
location = {Lisbon, Portugal},
series = {MSR '24}
}

@inproceedings{10.1145/3524610.3527911,
author={Tao, Chenning and Zhan, Qi and Hu, Xing and Xia, Xin},
  booktitle={2022 IEEE/ACM 30th International Conference on Program Comprehension (ICPC)}, 
  title={C4: Contrastive Cross-Language Code Clone Detection}, 
  year={2022},
  volume={},
  number={},
  pages={413-424},
  doi={10.1145/3524610.3527911},
  url={https://doi.org/10.1145/3524610.3527911}}

@article{scott1974cluster,
  title={ScottKnott: A Package for Performing the Scott-Knott Clustering Algorithm in R}, 
  volume={15}, 
  url={https://doi.org/10.5540/tema.2014.015.01.0003}, 
  DOI={10.5540/tema.2014.015.01.0003}, number={1}, 
  journal={Trends in Computational and Applied Mathematics}, 
  author={Jelihovschi, Enio and Faria, José Cláudio and Allaman, Ivan Bezerra}, 
  year={2014}, 
  month={Mar.}, 
  pages={003–017} 
  }

@book{montgomery2021introduction,
  author = {Montgomery, Douglas C. and Peck, Elizabeth A. and Vining, G. Geoffrey},
title = {Introduction to Linear Regression Analysis, Solutions Manual (Wiley Series in Probability and Statistics)},
year = {2007},
isbn = {0470125063},
publisher = {Wiley-Interscience},
address = {USA}
}

@article{mannkendall1998,
title = {A modified Mann-Kendall trend test for autocorrelated data},
journal = {Journal of Hydrology},
volume = {204},
number = {1},
pages = {182-196},
year = {1998},
issn = {0022-1694},
doi = {https://doi.org/10.1016/S0022-1694(97)00125-X},
url = {https://www.sciencedirect.com/science/article/pii/S002216949700125X},
author = {Khaled H. Hamed and A. {Ramachandra Rao}}
}

@INPROCEEDINGS{10356405,
  author       = {Md. Saeed Siddik and
                  Cor{-}Paul Bezemer},
  editor       = {Leon Moonen and
                  Christian D. Newman and
                  Alessandra Gorla},
  title        = {Do Code Quality and Style Issues Differ Across (Non-)Machine Learning
                  Notebooks? Yes!},
  booktitle    = {23rd {IEEE} International Working Conference on Source Code Analysis
                  and Manipulation, {SCAM} 2023, Bogot{\'{a}}, Colombia, October
                  2-3, 2023},
  pages        = {72--83},
  publisher    = {{IEEE}},
  year         = {2023},
  url          = {https://doi.org/10.1109/SCAM59687.2023.00018},
  doi          = {10.1109/SCAM59687.2023.00018},
  bibsource    = {dblp computer science bibliography, https://dblp.org}
  }

@article{LENARDUZZI2021110806,
title = {Does code quality affect pull request acceptance? An empirical study},
journal = {Journal of Systems and Software},
volume = {171},
pages = {110806},
year = {2021},
issn = {0164-1212},
doi = {https://doi.org/10.1016/j.jss.2020.110806},
url = {https://doi.org/10.1016/j.jss.2020.110806
},
author = {Valentina Lenarduzzi and Vili Nikkola and Nyyti Saarimäki and Davide Taibi}
}

@INPROCEEDINGS{8816735,
  author={Ferreira Campos, Uriel and Smethurst, Guilherme and Moraes, João Pedro and Bonifácio, Rodrigo and Pinto, Gustavo},
  booktitle={2019 IEEE/ACM 16th International Conference on Mining Software Repositories (MSR)}, 
  title={Mining Rule Violations in JavaScript Code Snippets}, 
  year={2019},
  volume={},
  number={},
  pages={195-199},
  doi={10.1109/MSR.2019.00039},
  url={https://doi.org/10.1109/MSR.2019.00039}}

@ARTICLE{9359361,
  author={Zhang, Haoxiang and Wang, Shaowei and Li, Heng and Chen, Tse-Hsun and Hassan, Ahmed E.},
  journal={IEEE Transactions on Software Engineering}, 
  title={A Study of C/C++ Code Weaknesses on Stack Overflow}, 
  year={2022},
  volume={48},
  number={7},
  pages={2359-2375},
  doi={10.1109/TSE.2021.3058985},
  url={https://doi.org/10.1109/TSE.2021.3058985}}

@INPROCEEDINGS{11025620,
  author = {Shin, Jiho and Tang, Clark and Mohati, Tahmineh and Nayebi, Maleknaz and Wang, Song and Hemmati, Hadi},
title = {Prompt Engineering or Fine-Tuning: An Empirical Assessment of LLMs for Code},
year = {2025},
publisher = {IEEE Press},
url = {https://doi.org/10.1109/MSR66628.2025.00082},
doi = {10.1109/MSR66628.2025.00082},
booktitle = {2025 IEEE/ACM 22nd International Conference on Mining Software Repositories (MSR)},
pages = {490–502},
numpages = {13},
location = {Ottawa, ON, Canada}
}

@misc{chatgpt2025,
  author = {{OpenAI}},
  title = {ChatGPT},
  howpublished = {\url{https://chatgpt.com/}},
  year = {2026},
  note = {Accessed: 2026-07-13}
}

@misc{claude2025,
  author = {{Anthropic}},
  title = {Claude AI: Next-generation AI assistant},
  howpublished = {\url{https://claude.ai/}},
  year = {2026},
  note = {Accessed: 2026-07-13}
}

@misc{openai2025privacy,
  author = {{OpenAI}},
  title = {Privacy Policy},
  howpublished = {\url{https://openai.com/policies/privacy-policy}},
  year = {2026},
  note = {Accessed: 2026-07-13}
}

@misc{anthropicPrivacy2025,
  author = {{Anthropic}},
  title = {Privacy Policy},
  howpublished = {\url{https://www.anthropic.com/legal/privacy}},
  year = {2026},
  note = {Accessed: 2026-07-13}
}

@article{grootendorst2022bertopic,
   title={BERTopic: Neural topic modeling with a class-based TF-IDF procedure}, 
      author={Maarten Grootendorst},
      year={2022},
      eprint={2203.05794},
      archivePrefix={arXiv},
      primaryClass={cs.CL},
      url={https://arxiv.org/abs/2203.05794}, 
}

@misc{sealXYZ2025,
  author = {{Software-Evolution-Analytics-Lab-SEAL}},
  title = {Replication Package},
  howpublished = {\url{https://github.com/Software-Evolution-Analytics-Lab-SEAL/CodeChat}},
  year = {2026},
  note = {Accessed: 2026-06-01}
}

@misc{codechat_dataset,
  author       = {Suzhen Zhong and Ying Zou and Bram Adams},
  title        = {{CodeChat V2.0}: Developer--LLM Conversations Dataset},
  howpublished = {\url{https://huggingface.co/datasets/Suzhen/CodeChat-V2.0}},
  year         = {2025},
  note         = {Accessed: 2026-08-27}
}

@misc{anthropic2025,
  author = {{Anthropic}},
  title = {Anthropic Economic Index: AI’s Impact on Software Development},
  howpublished = {\url{https://www.anthropic.com/research/impact-software-development}},
  year = {2026},
  note = {Accessed: 2026-07-13}
}

@misc{mdnDiv2025,
  author = {{Mozilla Contributors}},
  title = {The Content Division element},
  howpublished = {\url{https://developer.mozilla.org/en-US/docs/Web/HTML/Element/div}},
  year = {2026},
  note = {Accessed: 2026-07-13}
}

@misc{react2025,
  author = {{React Team}},
  title = {React - The library for web and native user interfaces},
  howpublished = {\url{https://react.dev}},
  year = {2026},
  note = {Accessed: 2026-07-13}
}

@inproceedings{paszke2019pytorch,
  author       = {Adam Paszke and
                  Sam Gross and
                  Francisco Massa and
                  Adam Lerer and
                  James Bradbury and
                  Gregory Chanan and
                  Trevor Killeen and
                  Zeming Lin and
                  Natalia Gimelshein and
                  Luca Antiga and
                  Alban Desmaison and
                  Andreas K{\"{o}}pf and
                  Edward Z. Yang and
                  Zachary DeVito and
                  Martin Raison and
                  Alykhan Tejani and
                  Sasank Chilamkurthy and
                  Benoit Steiner and
                  Lu Fang and
                  Junjie Bai and
                  Soumith Chintala},
  editor       = {Hanna M. Wallach and
                  Hugo Larochelle and
                  Alina Beygelzimer and
                  Florence d'Alch{\'{e}}{-}Buc and
                  Emily B. Fox and
                  Roman Garnett},
  title        = {PyTorch: An Imperative Style, High-Performance Deep Learning Library},
  booktitle    = {Advances in Neural Information Processing Systems 32: Annual Conference
                  on Neural Information Processing Systems 2019, NeurIPS 2019, December
                  8-14, 2019, Vancouver, BC, Canada},
  pages        = {8024--8035},
  year         = {2019},
  url          = {https://proceedings.neurips.cc/paper/2019/hash/bdbca288fee7f92f2bfa9f7012727740-Abstract.html},
  bibsource    = {dblp computer science bibliography, https://dblp.org}
}

@misc{chollet2015keras,
  author = {François Chollet},
  title = {Keras},
  year = {2015},
  publisher = {GitHub},
  howpublished = {\url{https://github.com/keras-team/keras}},
  note = {Accessed: 2026-07-13}
}

@misc{crawshaw2020multi,
  title={Multi-Task Learning with Deep Neural Networks: A Survey}, 
      author={Michael Crawshaw},
      year={2020},
      eprint={2009.09796},
      archivePrefix={arXiv},
      primaryClass={cs.LG},
      url={https://arxiv.org/abs/2009.09796}, 
}

@misc{pylint2025,
  author = {{Pylint Development Team}},
  title = {Pylint Tutorial and Documentation},
  howpublished = {\url{https://pylint.readthedocs.io/en/stable/tutorial.html}},
  year = {2026},
  note = {Accessed: 2026-07-13}
}

@misc{pmd2025,
  author = {{PMD Team}},
  title = {PMD - Source Code Analyzer Documentation},
  howpublished = {\url{https://pmd.github.io/pmd/index.html}},
  year = {2026},
  note = {Accessed: 2026-07-13}
}

@misc{eslint2025,
  author = {{ESLint Team}},
  title = {Getting Started with ESLint - Pluggable JavaScript Linter},
  howpublished = {\url{https://eslint.org/docs/latest/use/getting-started}},
  year = {2026},
  note = {Accessed: 2026-07-13}
}

@misc{cppcheck2025,
  author = {{Cppcheck Team}},
  title = {Cppcheck: A Static Analyzer for C/C++ Code},
  howpublished = {\url{https://cppcheck.sourceforge.io/}},
  year = {2026},
  note = {Accessed: 2026-07-13}
}

@misc{roslynAnalyzers2025,
  author = {{Microsoft}},
  title = {Roslyn Analyzers GitHub Repository},
  howpublished = {\url{https://github.com/dotnet/roslyn-analyzers}},
  year = {2026},
  note = {Accessed: 2026-07-13}
}

@misc{pep8,
  author = {Guido van Rossum and Barry Warsaw and Alyssa Coghlan},
  title = {PEP 8 -- Style Guide for Python Code},
  howpublished = {\url{https://peps.python.org/pep-0008}},
  year = {2026},
  note = {Accessed: 2026-07-13}
}

@Article{Kumar2025,
author={Kumar, Vinod
and Baghel, Anurag Singh},
title={BERTopic: a model to solve cold start problem in software bug recommendation system},
journal={International Journal of Information Technology},
year={2025},
month={Jun},
day={01},
volume={17},
number={5},
pages={3125-3130},
issn={2511-2112},
doi={10.1007/s41870-025-02470-8},
url={https://doi.org/10.1007/s41870-025-02470-8}
}

@Article{Gottumukkala2025,
author={Gottumukkala, Devi Priya
and P.V.G.D, Prasad Reddy
and Rao, S. Krishna},
title={Topic modeling-based prediction of software defects and root cause using BERTopic, and multioutput classifier},
journal={Scientific Reports},
year={2025},
month={Jul},
day={14},
volume={15},
number={1},
pages={25428},
issn={2045-2322},
doi={10.1038/s41598-025-11458-0},
url={https://doi.org/10.1038/s41598-025-11458-0}
}

@misc{zhong2026humanaisynergyagenticcode,
      title={Human-AI Synergy in Agentic Code Review}, 
      author={Suzhen Zhong and Shayan Noei and Ying Zou and Bram Adams},
      year={2026},
      eprint={2603.15911},
      archivePrefix={arXiv},
      primaryClass={cs.SE},
      url={https://arxiv.org/abs/2603.15911}, 
}

@inproceedings{csharp_humancode_so,
author = {Yang, Di and Hussain, Aftab and Lopes, Cristina Videira},
title = {From query to usable code: an analysis of stack overflow code snippets},
year = {2016},
isbn = {9781450341868},
publisher = {Association for Computing Machinery},
address = {New York, NY, USA},
url = {https://doi.org/10.1145/2901739.2901767},
doi = {10.1145/2901739.2901767},
booktitle = {Proceedings of the 13th International Conference on Mining Software Repositories},
pages = {391–402},
numpages = {12},
location = {Austin, Texas},
series = {MSR '16}
}

@Article{Barua2014,
author={Barua, Anton
and Thomas, Stephen W.
and Hassan, Ahmed E.},
title={What are developers talking about? An analysis of topics and trends in Stack Overflow},
journal={Empirical Software Engineering},
year={2014},
month={Jun},
day={01},
volume={19},
number={3},
pages={619-654},
issn={1573-7616},
doi={10.1007/s10664-012-9231-y},
url={https://doi.org/10.1007/s10664-012-9231-y}
}

@inproceedings{so_hci_humancode,
author = {Mamykina, Lena and Manoim, Bella and Mittal, Manas and Hripcsak, George and Hartmann, Bj\"{o}rn},
title = {Design lessons from the fastest q\&a site in the west},
year = {2011},
isbn = {9781450302289},
publisher = {Association for Computing Machinery},
address = {New York, NY, USA},
url = {https://doi.org/10.1145/1978942.1979366},
doi = {10.1145/1978942.1979366},
booktitle = {Proceedings of the SIGCHI Conference on Human Factors in Computing Systems},
pages = {2857–2866},
numpages = {10},
location = {Vancouver, BC, Canada},
series = {CHI '11}
}

@misc{li2025riseai,
      title={The Rise of AI Teammates in Software Engineering (SE) 3.0: How Autonomous Coding Agents Are Reshaping Software Engineering}, 
      author={Hao Li and Haoxiang Zhang and Ahmed E. Hassan},
      year={2025},
      eprint={2507.15003},
      archivePrefix={arXiv},
      primaryClass={cs.SE},
      url={https://arxiv.org/abs/2507.15003}, 
}

@misc{baumann2026swechat,
      title={SWE-chat: Coding Agent Interactions From Real Users in the Wild}, 
      author={Joachim Baumann and Vishakh Padmakumar and Xiang Li and John Yang and Diyi Yang and Sanmi Koyejo},
      year={2026},
      eprint={2604.20779},
      archivePrefix={arXiv},
      primaryClass={cs.AI},
      url={https://arxiv.org/abs/2604.20779}, 
}

@misc{zhong2026humancentricagenticcodereview,
      title={From Human-Centric to Agentic Code Review: The Impact of Different Generations of Generative AI Technology on Review Quality}, 
      author={Suzhen Zhong and Shayan Noei and Bram Adams and Ying Zou},
      year={2026},
      eprint={2607.13196},
      archivePrefix={arXiv},
      primaryClass={cs.SE},
      url={https://arxiv.org/abs/2607.13196}, 
}

\end{document}